\documentclass[twocolumn,aps,prx,showpacs,amsmath,superscriptaddress,longbibliography,notitlepage]{revtex4-1}
\usepackage{amssymb}
\usepackage{mathrsfs}
\usepackage{graphicx}
\usepackage{float}
\usepackage[caption=false]{subfig}
\usepackage[pdftex,colorlinks,linkcolor={red!75!black},citecolor={red!75!black},urlcolor={blue!75!black}]{hyperref}
\usepackage{verbatim}
\usepackage{epstopdf}
\usepackage[normalem]{ulem}
\usepackage{xcolor}
\usepackage{bm}
\usepackage{soul}
\usepackage{ulem}

\newcommand{\clr}{\color{red!75!black}}

\newcommand{\Rnum}[1]{\uppercase\expandafter{\romannumeral #1\relax}}

\def\LLH{\textcolor{cyan}}

\def\blue{\textcolor{blue}}

\begin{document}


\title{Dynamical suppression of many-body non-Hermitian skin effect in Anyonic systems}
\author{Yi Qin}
\affiliation{Guangdong Provincial Key Laboratory of Quantum Metrology and Sensing, School of Physics and Astronomy, Sun Yat-Sen University (Zhuhai Campus), Zhuhai 519082, China}
\author{Ching Hua Lee}\email{phylch@nus.edu.sg}
\affiliation{Department of Physics, National University of Singapore, Singapore 117542}
\author{Linhu Li}\email{rubilacxelee@gmail.com}
\affiliation{Guangdong Provincial Key Laboratory of Quantum Metrology and Sensing, School of Physics and Astronomy, Sun Yat-Sen University (Zhuhai Campus), Zhuhai 519082, China}
\affiliation{Quantum Science Center of Guangdong-Hong Kong-Macao Greater Bay Area (Guangdong), Shenzhen, China}

\begin{abstract}
The non-Hermitian skin effect (NHSE) is a fascinating phenomenon in nonequilibrium systems where eigenstates massively localize at the systems' boundaries, pumping (quasi-)particles loaded in these systems unidirectionally to the boundaries.
Its interplay with many-body effects have been vigorously studied recently, and inter-particle repulsion or Fermi degeneracy pressure have been shown to limit the boundary
accumulation induced by the NHSE both in their eigensolutions and dynamics. However, in this work we found that anyonic statistics can even more profoundly affect the NHSE dynamics, suppressing or even reversing the state dynamics against the localizing direction of the NHSE.
This phenomenon is found to be more pronounced when more particles are involved.
The spreading of quantum information in this system shows even more exotic phenomena, where NHSE affects only the information dynamics for a thermal ensemble, but not that for a single initial state.
Our results open up a new avenue on exploring novel non-Hermitian phenomena arisen from the interplay between NHSE and anyonic statistics, and can potentially be demonstrated in ultracold atomic quantum simulators and quantum computers.
\end{abstract}

\maketitle
\section*{Introduction} 
Picking up an arbitrary phase factor after exchange, anyons represent a more general calss of particles~\cite{Wilczek1982PRL,Tsui1982PRL,Laughlin1983PRL},
whose unusual statistics
induce many fascinating phenomena~\cite{Halperin1984PRL,Arovas1984PRL,Yao2007PRL,Bauer2014NC,kitaev2006A,Keilmann2011NC,Greschner2015PRL,Arcila2016PRA,Arcila2018PRA,Lange2017PRL,Liu2018PRL,Joyce2023arxiv}
and hold the promise to eventual fault-tolerant topological quantum computation and information processing~\cite{kitaev2003fault,DasSarma2005PRL,Nayak2008RMP,Carrega2021,iqbal2024non,lee2023par,LeeChing2015,LeeChing2018PRL}. Originally considered as two-dimensional quasiparticles, 1D anyonic statistics have also been predicted to emerge in cold bosonic atoms~\cite{Keilmann2011NC,Greschner2015PRL} and photonic systems~\cite{Yuan2017PRA,
Keilmann2011NC, Greschner2015PRL,Christoph2016PRL},
and have been emulated in circuit lattices by mapping their eigenmodes to two-anyon eigenstates~\cite{Zhang2022NC,Zhang2023CP}.
Assisted by Floquet engineering, arbitrary statistical phase of 1D anyons has been recently realized by Greiner's group~\cite{Joyce2023arxiv} in cold atom systems.

In the recent years, great attention has also been drawn towards another physical mechanism behind asymmetric dynamics, the non-Hermitian skin effect (NHSE)~\cite{Martinez2018,Yao2018}, which manifests as collosal accumulation of static
eigen-wavefunctions and states evolving over time
~\cite{ClaesPRB2021,manna2023CP,Zeng2020PRB,LeePRB2019,Okuma2020,Borgnia2020,Zhang2020PRL,Li2021,Liu2021,Roccati2021,li2022non,tai2023zoology,qin2023universal,zhang2022review,zhang2022review,lin2023topological,yang2022des,Jianghui2023,qin2023kin,lei2024ac,Aditi2024arxiv}.
~Entering the realm of many-body physics, novel extensions of NHSE have been uncovered during the past few years~\cite{Lee2021RS,Faugno2022PRL,Zheng2024PRL,Mu2022,Cao2023PRB,Louis2024SCP,MaoLiang2023,ZhangPRB2022,LeeC2021PRB,Xingran2021PRB,yoshida2023non,hamanaka2024,gliozzi2024many,Qin2024PRL}.
In particular, it has been found that NHSE can induce real-space Fermi surfaces for fermions and boundary condensation for bosons~\cite{Mu2022,Cao2023PRB,Louis2024SCP}, while the latter will be suppressed by a strong repulsive interaction~\cite{MaoLiang2023,Zheng2024PRL}.
In a recent study, 
an occupation-dependent NHSE is uncovered for hardcore bosons and fermions,  whose different exchange symmetries lead to distinguishable behaviors despite residing in the same Fock space~\cite{Qin2024PRL}.
On the other hand, the interplay between anyonic statistics and NHSE still remains largely unexplored.

In this paper, we report the discovery of a dynamical suppression of NHSE in a 1D non-Hermitian anyon-Hubbard model, revealing the intricate consequences of anyonic statistics acting on non-Hermitian physics.
Explicitly, we find that the dynamical evolution is not always in accordance with the static localization direction of eigenstates, which suffer from qualitatively the same NHSE at different statistical angles of the anyons. 
In particular, the state evolution may even experience a reversed density pumping process, 
during which the density evolves against the non-Hermitian pumping direction induced by NHSE.
Such a reversed pumping is found to be more pronounced when increasing the number of particles loaded in the system.
More drastically, by examining the out-of-time-ordered correlator (OTOC), 
we find that the information spreading is dominated by NHSE for a thermal ensemble,
but immune to NHSE for a single initial state at zero temperature.

\begin{figure*}[htbp]
\centering
\includegraphics[width=0.95\linewidth]{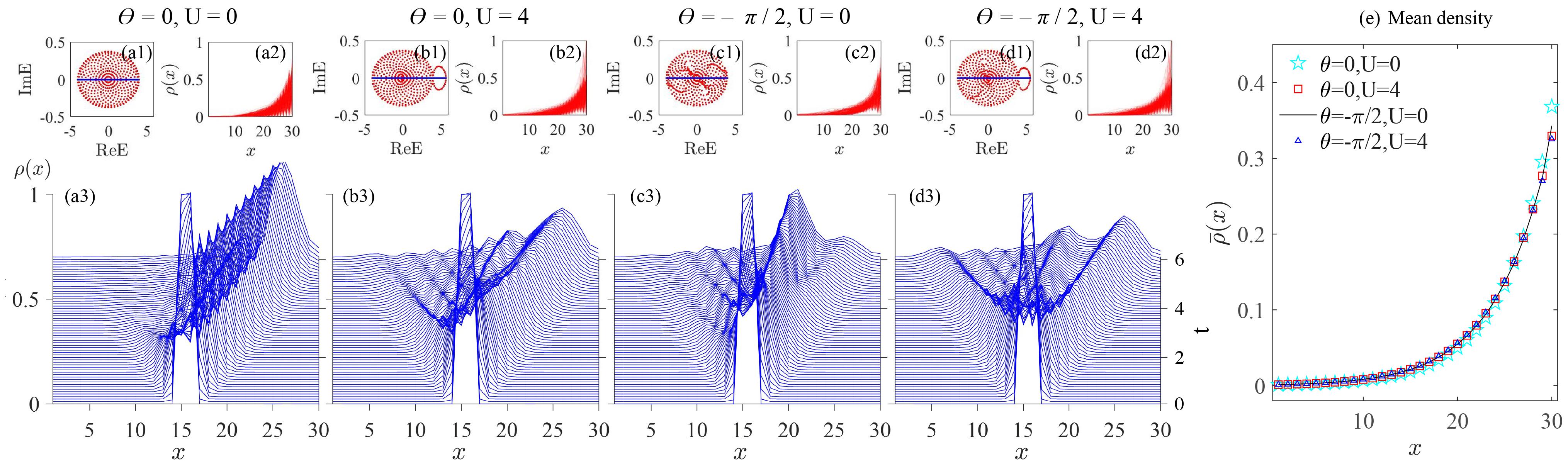}
\caption{\label{fig:DynamicsM}
Static properties and density evolutions for $N=2$ particles with different statistical angles $\theta$ and interaction strength $U$.
(a) Bosons ($\theta=0$) with zero interaction, with
(a1) eigenenergies of the model under PBCs (red) and OBCs (blue); (a2) particle distribution $\rho(x)$ of all many-body eigenstates (pink); and (a3) density evolution for two particles evenly distributed at the center of the 1D chain.
(b), (c), and (d) displayed the same quantities for systems with different statistical angles and interaction strengths, as labeled on top of each panel. (e) The almost identical average density $\bar{\rho}_{x}$ for all states in (a2) to (d2), represented by cyan star, red square, black line, and blue triangular, respectively. It is seen that anyonic statistics have little effect on the distribution of eigenstates, even though they cause distinguished dynamics.
Other parameters are $J_L=e^{-\alpha},J_R=e^{\alpha}, \alpha=0.1$ and $L=30$. 
In each of (b1) and (d1), some eigenenergies form a loop separated from the others, corresponding to two-particle bound states induced by the Hubbard interaction (see Supplemental Note 1).
}
\end{figure*}   

\section*{Results}
\subsection*{ NHSE in a 1D anyonic lattice}

We consider a one-dimensional non-Hermitian anyon-Hubbard  model (NHAHM) described by the Hamiltonian
\begin{equation}
{{\hat H}_A} =  - \sum\limits_{j = 1}^{L - 1} {\left( {{J_L}\hat a_j^\dag {{\hat a}_{j + 1}} + {J_R}\hat a_{j + 1}^\dag {{\hat a}_j}} \right)}  + \frac{U}{2}\sum\limits_{j = 1}^L {{{\hat n}_j}({{\hat n}_j} - 1)} ,
\end{equation}
where $J_L=e^{-\alpha}$ and $J_R=e^{\alpha}$ with $\alpha>0$ describe the non-Hermitian nearest-neighbor hopping amplitudes that induce NHSE,
$U$ is the onsite Hubbard interaction,
and ${{\hat n}_j} = \hat a_j^\dag {{\hat a}_j}$. The communication relations are obeyed by the anyonic creation ($ \hat a_j^\dag$) and annihilation ($\hat a_j$) operators,
\begin{equation}
\begin{array}{l}
{[{{\hat a}_j},{{\hat a}_k}]_\theta } \equiv {{\hat a}_j}{{\hat a}_k} - {e^{ - i\theta {\mathop{\rm sgn}} (j - k)}}{{\hat a}_k}{{\hat a}_j} = 0,\\
{[{{\hat a}_j},\hat a_k^\dag ]_{ - \theta }} \equiv {{\hat a}_j}\hat a_k^\dag  - {e^{i\theta {\mathop{\rm sgn}} (j - k)}}\hat a_k^\dag {{\hat a}_j} = {\delta _{jk}},\label{commute}
\end{array}
\end{equation}
where ${\rm sgn}(x)$ is the sign function
and $\theta$ is the statistical angle. 
$\theta=0$ and $\theta=\pi$ represent normal bosons and ``pseudofermions" that obey bosonic statistic only when occupying the same lattice site, respectively~\cite{Keilmann2011NC}.
Via a generalized Jordan-Wigner transformation ${\hat a_j} = {\hat b_j}{e^{ - i\theta \sum\limits_{k = 1}^{j - 1} {\hat n_k^{}} }}$,
the anyonic model can be mapped to an extended Bose-Hubbard model with a density-dependent phase factor acquired by particles hopping between sites,
which facilitates further analysis. 
Under this mapping, the anyonic Hamiltonian $\hat H_A$ is mapped to
\begin{equation}\label{EBHM}
\begin{array}{lllll}
{{\hat H}_B} =  &  - \sum\limits_{j = 1}^{L - 1} {\left( {{J_L}\hat b_j^\dag {e^{ - i\theta {{\hat n}_j}}}{{\hat b}_{j + 1}} + {J_R}\hat b_{j + 1}^\dag {e^{i\theta {{\hat n}_j}}}{{\hat b}_j}} \right)} \\
 &  + \frac{U}{2}\sum\limits_{j = 1}^L {{{\hat n}_j}({{\hat n}_j} - 1)},
\end{array}
\end{equation}
where $ \hat b_j^\dag$ ($ \hat b_j$) is bosonic creation (annihilation) operators and ${{\hat n}_j} = \hat a_j^\dag {{\hat a}_j}=\hat b_j^\dag {{\hat b}_j}$. 
Using Floquet engineering,
similar density-dependent terms giving rise to anyonic statistics have recently been realized in ultracold $^{87}$Rb atoms~\cite{Joyce2023arxiv},
and the asymmetric hopping amplitudes $J_{L}$ and $J_R$ may be implemented with site-dependent atomic loss induced by near-resonant light with position-dependent intensity~\cite{Faugno2022PRL,takasu2020pt},
making it possible to realize our model in cold atom systems.

Diagonalizing the Hamiltonian $\hat{H}_B$, we confirm that complex eigenenergies and NHSE arise in this model due to the asymmetric hopping, as shown in the top row of Fig.~\ref{fig:DynamicsM}. The divergence between PBC and OBC spectra indicates the emergence of NHSE under OBCs, as evidenced by the massive accumulation of eigenstates in Fig. \ref{fig:DynamicsM}(a2) to (d2).
The NHSE can be further characterized by a spectral winding number in terms of a $U(1)$ gauge field~\cite{Zhang2020PRL, Okuma2020, Borgnia2020, Kawabata2022PRB}, as demonstrated in the Supplemental Note 3.
A key observation is that the spatial distribution for all eigenstates and their average are seen to be roughly the same under different statistical angle $\theta$ and the interaction strength $U$ [Fig. \ref{fig:DynamicsM}(e)], implying that the anyonic statistics have little effect on the NHSE at the static level.
\blue{Physically, this is because the statistic angle does not affect the left and right hopping magnitudes, whose difference gives raise to the NHSE, as further elaborate in the Methods section.}
\blue{It should be emphasized that we have chosen $\theta = -\pi/2$ and $0$ in Fig. \ref{fig:DynamicsM} to compare anyonic and bosonic scenarios, and other values of $\theta$ also produce qualitatively the same results (no shown).
In particular, to have a better demonstration between the interplay between NHSE and anyonic properties,
$\theta = -\pi/2$ is chosen as it corresponds to the strongest asymmetric anyonic dynamics to the left~\cite{Liu2018PRL} (also see later discussion), 
opposite to the non-Hermitian pumping for the parameters we consider.}



\subsection*{Dynamical suppression of NHSE}
It is commonly assumed that the localizing direction of NHSE indicates the tendency of the state dynamics governed by the non-Hermitian Hamiltonian~\cite{Qin2024PRL}.
However, despite the nearly identical behavior of NHSE in our model, 
we find that the dynamics depends strongly on the interaction strength and statistic angle, 
and may even violate the prediction of NHSE to a certain extent. 
We consider the density evolution for anyons uniformly distributed at the center of a chain with one particle per site,
with the initial state given by ${\left| {{\Psi^A_0}} \right\rangle} = \prod\nolimits_i {\hat a_i^\dag \left| 0 \right\rangle}$.
With the generalized Jordan-Wigner transformation, the time-dependent density distribution of anyons can be expressed as
 \begin{eqnarray}\label{DenAB}
n_j^A(t)
&=& \left\langle {{\Psi^A_0}} \right|{e^{i{{\hat H}_A}^{\dag}t}}\hat n_j{e^{-i{{\hat H}_A}t}}{\left| {{\Psi^A_0}} \right\rangle}\nonumber\\
&=& \left\langle {{\Psi^B_0}} \right|{e^{i{{\hat H}_B}^{\dag}t}}\hat n_j{e^{-i{{\hat H}_B}t}}{\left| {{\Psi^B_0}} \right\rangle} =  n_j^B(t), 
 \end{eqnarray}
thus the anyon dynamics can be directly measured from the mapped bosonic density $n_j^B(t)$. 
We note that in our model, the time-dependent density satisfies ${\left\langle {\hat n_j^B(t)} \right\rangle _{ + U,\theta }} = {\left\langle {\hat n_j^B(t)} \right\rangle _{ - U, - \theta }}$, \blue{protected by 
a combined symmetry
${{\cal K}}{\hat H_B}{{{\cal K}}^\dag } = \hat H_B^\dag$ of the Hamiltonian with ${{\cal K}} = {{{\cal R}}_z}{{\cal I}{\cal T}}$,
where ${\cal I}$, ${\cal T}$, and ${{{\cal R}}_z} = {e^{ - i\theta {{\hat n}_j}({{\hat n}_j} - 1)/2}}$ 
represent the operators for inversion symmetry, time-reversal symmetry, and a number-dependent gauge transformation, respectively (see Methods). That is, cases with repulsive and attractive interactions can be mapped to each other by reversing the sign of $\theta$.} \blue{Thus
we shall focus only on the case with $U\geqslant0$ and $\theta$ ranging from $-\pi$ to $\pi$ without loss of generality.}

\blue{Following, we focus on how a state evolves under a non-Hermitian Hamiltonian.
Specifically, for an initial state $|\psi^A_0\rangle$,
we normalize the final state at time $t$ as
\begin{equation}
\left|\psi(t)\right\rangle=\frac{e^{-i \hat{H} t / \hbar}\left|\psi_0^A\right\rangle}{\sqrt{\left\langle\psi_0^A\left|e^{i \hat{H}^{\dagger} t / \hbar} e^{-i \hat{H} t / \hbar}\right| \psi_0^A\right\rangle}}.
\end{equation}
}
The density evolutions for $N=2$ anyons with different interaction strengths and statistical angles are shown in bottom panels of Fig.~\ref{fig:DynamicsM}. 
Unbalanced pumping induced by NHSE can be most clearly seen in Fig.~\ref{fig:DynamicsM}(a3) with $\theta=0$ and $U=0$, 
where the particle density shows a unidirectional ballistic evolution toward the right.
A finite interaction is known to suppress the expansion of bosons and lead to a diffusive dynamics~~\cite{Ronzheimer2013PRL}, thus weakening the unidirectional evolution, as can be seen in Fig.~\ref{fig:DynamicsM}(b3).

Away from the bosonic limit at $\theta=0$, 
the anyonic statistics induce an asymmetric particle transport~\cite{Liu2018PRL,Wang2022PRB,Joyce2023arxiv} \blue{(also see Methods)}.
It can further suppress the NHSE-induced right-moving tendency,
and the dynamics show signatures more of a diffusive evolution instead of a ballistic one,
as can be seen in Fig.~\ref{fig:DynamicsM}(c3) and (d3) for $\theta=-\pi/2$.
Note that the seemingly ballistic evolution with a smaller velocity in Fig.~\ref{fig:DynamicsM}(c3) is an exception only for $N=2$ particles,
and becomes diffusive when the particle number increases, as shown in Supplemental Note 2 D. 
The most peculiar thing is that upon turning on the interaction, 
the asymmetric transport of anyons may even overwhelm the NHSE at the beginning of the evolution,
resulting in an evolution opposite to the direction of skin localization for a short period of time, as shown in Fig.~\ref{fig:DynamicsM}(d3).

\begin{figure}[htbp]
\centering
\includegraphics[height=8.5cm]{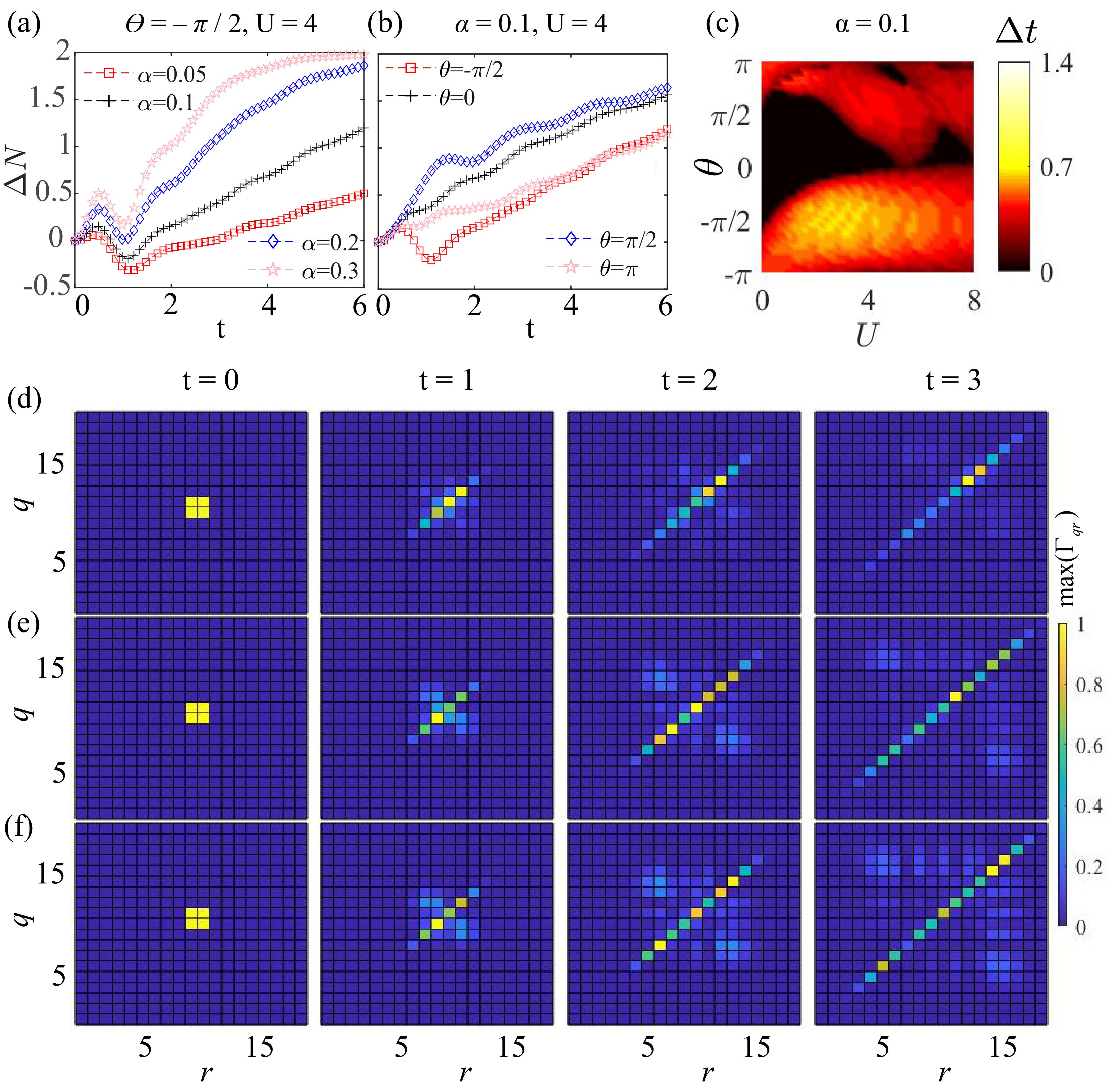}
\caption{\label{fig_AlphaChange} Reversed pumping and density-density correlation for $N=2$.
(a) The density imbalance $\Delta N$ for $\theta=-\pi/2$ with $U=4$, for different non-Hermitian amplitudes with $\alpha=0.05,0.1,0.2,0.3$ respectively. 
(b) The same density imbalance $\Delta N$ for various $\theta$ with $\alpha=0.1$, $U=4$.  
(c) A phase diagram demonstrating the reversed-pumping time $\Delta t$, defined as the interval with $\frac{d\Delta N}{dt}<0$. 
Nonzero $\Delta t$ is also seen around $\theta\in(0,\pi)$ with relatively large $U$, which is resulted from the fluctuation of $\Delta N$ at larger $t$ induced by the interference dynamics in Fig. \ref{fig:DynamicsM} (see Supplemental Note 2 A).
(d) - (f)
Density-density correlation of the evolved state at different time $t$, with $\theta=-\pi/2$, $\alpha=0.1$ and $L=20$, and $U=0,4,8$ from (d) to (f) respectively. As $U$ increases, the diagonal spreading of the correlation shows a bidirectional pumping toward both $q=r=L$ and $q=r=1$, indicating the NHSE and the reversed density pumping induced by anyonic statistics, respectively.
}
 \end{figure}

\subsection*{Reversed density pumping}

To characterize the competition between NHSE and statistics-induced asymmetric dynamics, we study the time-dependent density imbalance $\Delta N = \sum\nolimits_{i = 1}^{L/2} {\left( {{n_{i + L/2}} - {n_i}} \right)} $ between the two halves of the 1D chain.
As shown in Fig. \ref{fig_AlphaChange}(a) for $N=2$, at $\theta=-\pi/2$ and $U=4$,
the anyonic density evolution shows a reversed pumping against the NHSE,
with $\Delta N$ decreases with time $t$ when $0.5\lesssim t\lesssim1$. 
Such a reversed pumping is seen to be robust even under relatively strong non-Hermitian pumping strengths, e.g., $\alpha=0.3$ in the figure, where the density imbalance always favours the direction of NHSE ($\Delta N>0$) and saturated to $\Delta N=2$ rapidly. 
\blue{It should be noted that for a much weaker non-Hermiticity, the pumping of NHSE will be fully suppressed (with $\Delta N<0$) during a longer period of time, as shown in the Supplement Note 2 C.}
For comparison, we also plot $\Delta N$ versus time  for $\alpha=0.1$ and $U=1$ at different statistical angles in Fig.~\ref{fig_AlphaChange}(b). 
For the several chosen values of $\theta$,
the reversed pumping process \blue{with decreasing $\Delta N$} can be clearly identified only when $\theta=-\pi/2$.
Furthermore, $\Delta N$ is seen to increase faster for $\theta\neq -\pi/2$, showing a domination of NHSE on the state dynamics.
To characterize the magnitude of reversed pumping, we further consider its duration $\Delta t$ as an indicator, defined as the time interval where $\frac{d\Delta N}{dt}<0$.
As shown in Fig. \ref{fig_AlphaChange}(c), $\Delta t$ reaches it maximum at $\theta\approx -\pi/2$ and $U\approx 3$. 
We note that such a reversed density pumping relies crucially on the anyonic statistics,
and may disappear if the particles initially occupy the same lattice site (acting as bosons with $\theta=0$), or are separated from each other by at least one site (acting as single particles), as shown in  Supplemental Note 2 B.
In addition, $\Delta t$ also takes small but nonzero values for $\theta\in[0,\pi]$. This is because the diffusive anyonic dynamics causes interference between different portions of the evolving state,
 resulting in certain fluctuation of $\Delta N$ that shows weak reversed pumping, as can be seen from the data for $\theta=\pi/2$ in Fig. \ref{fig_AlphaChange}(b).

To provide a full picture of the different diffusive and unidirectional dynamics in the system,
we calculate the density-density correlation defined as
\begin{equation}\label{TwoP}
\Gamma_{qr}=\langle \psi (t) |\hat{n}_q\hat{n}_r |\psi(t)\rangle,
\end{equation}
and display the results for two particles with $\theta=-\pi/2$ and different values of $U$ in Fig.~\ref{fig_AlphaChange}(d) to (f).When $U=0$, the dynamics mainly reflects the unidirectional pumping of NHSE, 
as nonzero $\Gamma_{qr}$ mostly distributes along diagonal ($q=r$), and its peak moves toward $q=r=L$ during the evolution [Fig.~\ref{fig_AlphaChange}(d)].
Turning on the interaction, we can see in Fig.~\ref{fig_AlphaChange}(e) and (f) that nonzero off-diagonal correlations appear with the distance between the two position ($|q-r|$) increases with time, indicating the diffusion enhanced by interaction.
On the other hand, a second peak of diagonal correlation appears and move toward $q=r=1$, signals the reversed density pumping caused by anyonic statistics.
The above discussion is focused on $N=2$ and we stress that it also holds for larger $N$, as shown in Supplemental Note 4.

\begin{figure}[htbp]
\centering
\includegraphics[width=1\linewidth]{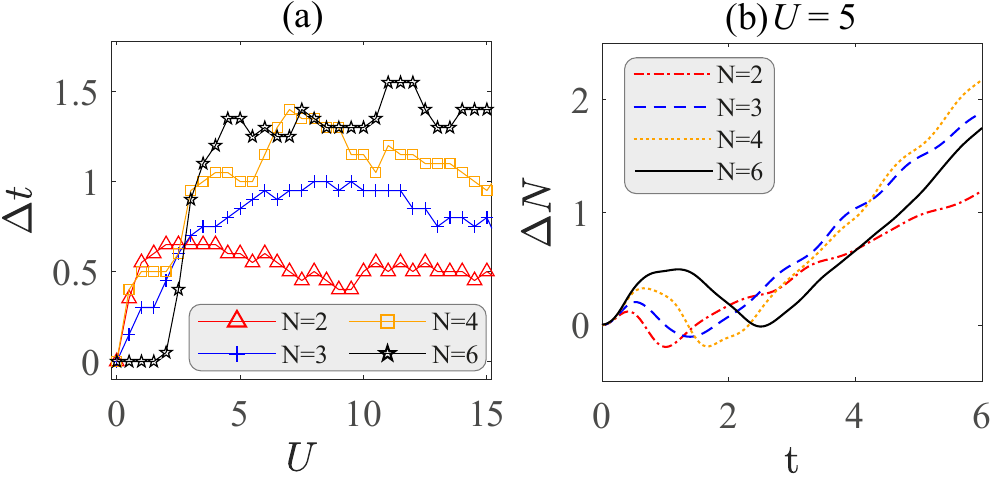}
\caption{\label{fig:N2346} Reversed pumping for various particle numbers for $\theta=-\pi/2$. (a) Reversed pumping time $\Delta t$ for $N=2,3,4,6$, represented by red triangles, blue pluses, gold squares, and black stars, respectively. 
(b) Density imbalance at $U=5$ with $N=2,3,4,6$, represented by red dotted line, blue dashed line, gold dot line, and black solid line, respectively.  
$\alpha=0.1$ is chosen for both panels.
The system's size is chosen to be $L=30$ for $N=2,4,6$, and $L=31$ for $N=3$. In the latter case, the density at the center of the system ($j=16$) is excluded when calculating $\Delta N$. \blue{We calculate the data for $N=6$ via the open-source tenpypackage.}
}
\end{figure}   

\subsection*{Reversed pumping with larger particle numbers $N$}

As our model contains only nearest-neighbor hopping, the anyonic statistics can function normally only for adjacent particles.
 Therefore, the reversed pumping it induces is expected to become more prominent with more particles in the system, distributing next to each other initially.
In Fig.~\ref{fig:N2346} (a), we demonstrate the reversed pumping time $\Delta t$ for $N=2,3,4,6$ of $\theta=-\pi/2$ with various interaction $U$. 
For $N=2$, $\Delta t$ is increasing fast around 0.5 and reaches its maximum approximately at $U=3$, then it shows a slightly decreasing behavior. While for $N=3,4$ and $6$, $\Delta t$ is increasing with the increase of $U$ (also see Supplemental Note 6). Interestingly, for $U\gtrsim 3$, the reversed pumping effect shows an interaction enhanced tendency,
as can be clearly seen from the Fig.~\ref{fig:N2346} (b) for $U=5$. 
Note that the slope of $N=2$ is seen to be smaller than the others for $t\gtrsim 2$. 
This is because with fewer particles, it takes shorter time for their wave-function to be mostly pumped to the right half of the lattice by NHSE, after which ($t\approx 2$ for $N=2$) $d \Delta N/d t$ becomes smaller as the remaining density on the left becomes negligible. 
For $N>2$,  the slope of $\Delta N$ also decreases similarly at larger $t$, as shown in Supplemental Note 2 E.
In addition, 
\blue{weak damped fluctuation} of $\Delta N$ is seen even when the dynamics is dominated by NHSE at larger $t$,
\blue{resultant from the interplay between anyonic interference and density diffusion. Namely, while the interference of anyons induces the fluctuation, the diffusion reduces local density, weakening the density being pumped across the center and resulting in an overall damping effect.} 

\subsection*{Out-of-time correlator and information spreading}
\begin{figure}[htbp]
\centering
\includegraphics[width=1\linewidth]{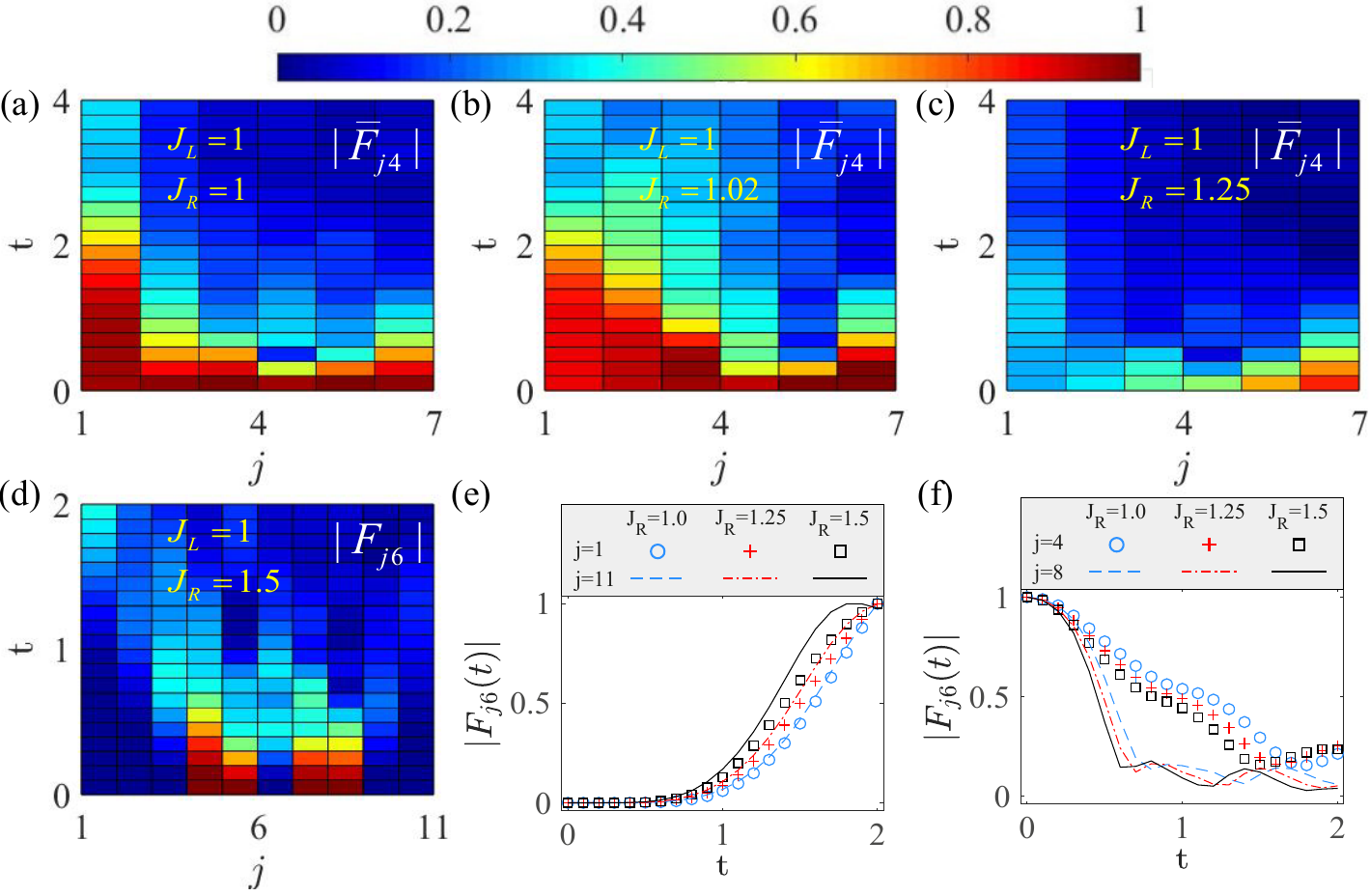}
\caption{\label{fig:OTOC} Information spreading for $\theta=-\pi/2,U=4$. (a) to (c) OTOC for a thermal ensemble at finite temperature with (a) $J_R=1.0$, (b)$J_R=1.02$, and (c) $J_R=1.25$. The system has $L=7$ lattice sites and  $N=4$ particles, with the inverse temperature $\beta=1/6$. The information spreading tends to move towards the right side with the increasing of $J_R$. 
(d) to (f) OTOC for a single state with $L=11,N=5$. For a single state, the information spreading hardly changes with the increase of $J_R$.
(d) $|F_{j6}|$ for $L=11$,$U=4$, and $J_R=1.5$. 
The initial state is chosen as $\psi(0)\rangle = |\hat{a}_4\hat{a}_5\hat{a}_6\hat{a}_7\hat{a}_8\rangle$. 
(e) $|F_{1,6}|$ and $|F_{11,6}|$ for $J_R=1,1.25,1.5$, respectively. 
(f) $|F_{4,6}|$ and $|F_{8,6}|$ for $J_R=1,1.25,1.5$, respectively.  $J_L=1$ is set in all panels. 
Results are normalized by setting $\max_{j,t}|\bar F_{j,4}|=1$ in (a) to (c), 
$\max_{t}|F_{j,6}|=1$ in (d),
and $\max_{t}|F_{j,6}|=1$ in (e) and (f).
}
\end{figure}

Having unveiled the sophisticated evolution of particle density in non-Hermitian anyonic systems.
it is natural to ask
how the anyonic statistics and NHSE simultaneously affect the dynamics of other physical quantities, such as the spreading of quantum information that can be characterized by the OTOC in non-Hermitian systems ~\cite{OTOCPRB2020}.
To describe the information spreading,
we first consider the OTOC of anyons for an ensemble defined as (see Supplemental Note 5 A for more details). 
\begin{eqnarray}
{C_{jk}}(t) = {\left\langle {{{\left| {{{\left[ {{{\hat a}_j}(t),{{\hat a}_k}(0)} \right]}_\theta }} \right|}^2}} \right\rangle _\beta },\label{C_ensemble}
\end{eqnarray}
where $\beta$ is the inverse temperature and ${\left\langle {\hat{O}} \right\rangle _\beta }$ means the thermal ensemble average ${\rm Tr}\left(e^{ - \beta {{\hat H}_A}}\hat{O}\right)/{\rm{Tr}}\left( {{e^{ - \beta {{\hat H}_A}}}} \right)$ of an operator $\hat{O}$.
$C_{jk}(t)$ describes the information propagated from site $k$ to site $j$ at time $t$, and ${C_{jk}}(0)=0$ is ensured by the generalized commutation relations of Eqs. \ref{commute},
The out-of-time-ordered part of the commutator is then given by~\cite{Swingle2016,Shen2017PRB}
\begin{eqnarray}
\bar{F}_{jk}(t)=\left\langle \hat{a}_j^\dagger(t)\hat{a}_k^\dagger(0)\hat{a}_j(t)\hat{a}_k(0)\right\rangle_\beta{e^{i\theta {\mathop{\rm sgn}} (j - k)}}.\label{F_ensemble}
\end{eqnarray}
\blue{The numerical results with $\theta=-\pi/2$ and $U=4$, i.e., under the parameters where the reversed density pumping strength nearly reaches its maximum, in Fig. \ref{fig:OTOC}(a) to (c).
It is seen that 
the direction of OTOC reverses from left to right when increasing the non-Hermiticity,
reflecting the competition between NHSE and the statistic-induced asymmetric OTOC spreading.
This is in contrast to bosonic systems without the anyonic asymmetric OTOC spreading, 
where the NHSE-induced asymmetric OTOC spreading can be observed even with a very weak non-Hermiticity, as shown in Supplemental Note 5 B.}
Physically, this is because the ensemble average represents a linear combination of eigenstates with different powers, which are all skin-localized toward the right when $J_R>J_L$ in our model.

In contrast, we find qualitatively different behaviors of the OTOC for a single initial state, whose definition is similar to Eqs. \eqref{C_ensemble} and \eqref{F_ensemble} but with $\beta=0$ and the ensemble average replaced by the average on the state.
In Fig. 4(d), we find that the information spreading for a single initial state [$F_{jk}(t)$] with a uniform distribution at the center of the 1D chain shows a clear asymmetric distribution with large value of OTOC at the left side and nearly unchanged even with a strong non-Hermiticy, reflecting the statistic-induced asymmetric OTOC spreading and immunity to NHSE.

In Fig. \ref{fig:OTOC}(e) and (f), we illustrate the information propagated from the center ($k=6$) to a few different positions $j$, with different non-Hermitian parameters.
As the initial state does not occupy the two ends of the system (with $11$ lattice sites),
we can see that $F_{j6}$ vanishes for $j=1$ or $11$ at small $t$, and increases monotonically with time.
On the other hand, we have $F_{j6}=1$ at the beginning and decreases with time for $j=4$ and $8$, which are the lattice sites occupied by the initial state. Nonetheless, these trends of OTOC are found to remain the same for both Hermitian ($J_R=1$) and non-Hermitian cases ($J_R\neq 1$), further verifying the dissimilar behaviors of OTOC for an ensemble at finite temperature (dominated by NHSE) and for a single initial state (immune to NHSE).

\section*{Conclusions}

We have revealed a dynamical suppression of NHSE by anyonic statistics, where the density evolutions show different diffusive or reversed pumping dynamics at different statistic angles.
In recent literature, it has been shown that NHSE can be suppressed through various means, e.g., by introducing magnetic ~\cite{OkumaPRL2019,LuMingPRL2021}
or electric fields~\cite{PengPRB2022}
with both the static solutions and dynamical evolutions changed drastically from ones of the NHSE in the suppression phase.

However, our results show that the anyonic statistics will affect only the density dynamics, whereas static solutions still manifest the same properties of NHSE under different statistical angles.
The reversed pumping process is shown to have a longer duration with larger numbers of particles, indicating that it may be easier for observation in the thermodynamic limit.
The coexistence between different diffusive dynamics, non-Hermitian pumping of NHSE, and reversed pumping are further demonstrated by the density-density correlation of the evolved state.
Finally, we also calculate the OTOC to characterize the quantum information spreading, 
which is found to be governed by NHSE only for a thermal ensemble.
On the other hand, OTOC calculated for a single initial state curiously follows that that of Hermitian limit dynamics, regardless of the strength of non-reciprocal pumping induced by non-Hermiticity.
These observations challenge the correspondence between static NHSE and unidirectional state dynamics, 
which is commonly assumed to be true in most theoretical~\cite{lee2019topological,song2019nonH,Qin2024PRL,PRLLi2020,LiPRB2022,zhangKai2022NC,kawabata2023entanglement}
 and experimental investigation~\cite{xiaoLei2020,Sebastian2020,Palacios2021,LiangPRL2022,gao2022anomalous,gu2022transient,zhang2023electrical,shen2023Ob}
~in the NHSE, particularly in quantum simulators~\cite{LiangPRL2022,shen2023Ob}.
Following this path, we may expect even more sophisticated non-Hermitian phenomena to arise from the interplay between anyonic statistics and other novel single-particle dynamics induced by NHSE, such as the non-Hermitian edge burst~\cite{XuePRL2022,WenPRA2024}
and self-healing of skin modes~\cite{Longhi2022PRL}.  
\blue{Another potential direction is to investigate non-Hermitian anyonic dynamics in the hardcore limit, 
which has been found to induce
an occupation-dependent particle separation for bosons and fermions~\cite{Qin2024PRL}. Nonetheless, long-range couplings are required for our model to manifest anyonic features in the hardcore limit (see Supplemental Note 7 for results or our model in the hardcore limit).}
\blue{Finally, we note that in this paper we only address Abelian anyons in a 1D lattice.
Beyond this simple picture, non-Abelian anyons may also emerge as quasiparticlesin other higher-dimensional systems, such as fractional quantum Hall systems and quantum spin liquids, where anyonic porperties may affect NHSE in different manners.
For example, it has been demonstrated that non-Abelian fusion of Fibonacci anyons induces exotic Bloch oscillations~\cite{Zhou2024PRB}, whose interplay with non-Hermitian pumping may lead to distinct dynamical behaviors that await further investigation.
}


\section*{Methods}
\subsection*{
Static NHSE and its independence from anyonic statistics
}
\blue{The origin of NHSE can be understood from a similarity transformation of the Hamiltonian matrix.
for the OBC Hamiltonian
\begin{eqnarray}
\hat{H}_B&=&-\sum_{j=1}^{L-1}{\left( J_L\hat{b}_{j}^{\dagger}e^{-i\theta \hat{n}_j}\hat{b}_{j+1}+J_R\hat{b}_{j+1}^{\dagger}e^{i\theta \hat{n}_j}\hat{b}_j \right)}\nonumber\\
&&+\frac{U}{2}\sum_{j=1}^L{\hat{n}_j(\hat{n}_j-1)},
\end{eqnarray}
we introduce a set of operators defined as 
\begin{eqnarray}
\hat{g}_{l}^{\dagger}=e^{-l\alpha}\hat{b}_{l}^{\dagger},\hat{g}_l=e^{l\alpha}\hat{b}_l.\label{eq:mapping}
\end{eqnarray} 
Then the above Hamiltonian can be rewritten as 
\begin{eqnarray}
\hat{H}_B&=&-\sum_{j=1}^{L-1}{\left( \hat{g}_{j}^{\dagger}e^{-i\theta \hat{n}_j}\hat{g}_{j+1}+\hat{g}_{j+1}^{\dagger}e^{i\theta \hat{n}_j}\hat{g}_j \right)}\nonumber
\\&&+\frac{U}{2}\sum_{j=1}^L{\hat{n}_j(\hat{n}_j-1)}, \hat{n}_j=\hat{g}_{j}^{\dagger}\hat{g}_j.
\end{eqnarray}
This transformation results in a Hermitian Hamiltonian matrix $H'_{B}$ given by $H'_{B}=SH_BS^{-1}$,
with $H_B$ the matrix form of $\hat{H}_B$ in the Fock basis of $\sum_l\hat{b}_l^\dagger|0\rangle$,
$H'_B=(H'_B)^\dagger$ the matrix form of $\hat{H}_B$ in the new Fock basis of $\sum_l\hat{g}_l^\dagger|0\rangle$,
and $S$ the transformation matrix describing the relation between operators $\hat{b}_l$ and $\hat{g}_l$.}

\blue{Note that by definition, $S$ is a similarity transformation that keeps only the eigenvalues unchanged. 
On the other hand,
the eigenvectors in the original basis can be obtained as
$\psi=S^{-1}\psi'$, with $\psi'$ an eigenvector of $H'_B$.
As $H'_B$ is a Hermitian matrix describing a system with translational symmetry within the bulk,
$\psi'$ generally describes extended states with uniform distributions of particles in the bulk.
However, the mapping between operators in Eq.~\eqref{eq:mapping} indicates that the similarity transformation of $S$ is not uniform, where a particle created at site $l$ has a position-dependent magnitude $e^{-l\alpha}$. Thus, for a positive $\alpha$,
an extended state of $\psi'$ with uniform distribution of particles becomes a boundary-localized state with amplified magnitude toward larger $l$.
Finally, as the mapping of Eq.~\eqref{eq:mapping} is irrelevant to the statistic angle $\theta$, we can conclude that the anyonic statistic do not affect the localizing direction of static NHSE.
}


\subsection*{Pseudo-Hermitian symmetry of the static Hamiltonian}
The Hamiltonian of the non-Hermitian anyon-Hubbard model is given by
\begin{equation}
{{\hat H}_A} =  - \sum\limits_{j = 1}^{L - 1} {\left( {{J_L}\hat a_j^\dag {{\hat a}_{j + 1}} + {J_R}\hat a_{j + 1}^\dag {{\hat a}_j}} \right)}  + \frac{U}{2}\sum\limits_{j = 1}^L {{{\hat n}_j}({{\hat n}_j} - 1)}.
\end{equation}
With a generalized Jordan-Wigner transformation ${\hat a_j} = {\hat b_j}{e^{ - i\theta \sum\limits_{k = 1}^{j - 1} {\hat n_k^{}} }}$,
the model is mapped to a boson-Hubbard model with a density-dependent phase factor,
\begin{equation}
\begin{array}{lllll}
{{\hat H}_B} =  &- \sum\limits_{j = 1}^{L - 1} {\left( {{J_L}\hat b_j^\dag {e^{ - i\theta {{\hat n}_j}}}{{\hat b}_{j + 1}} + {J_R}\hat b_{j + 1}^\dag {e^{i\theta {{\hat n}_j}}}{{\hat b}_j}} \right)} \\
& + \frac{U}{2}\sum\limits_{j = 1}^L {{{\hat n}_j}({{\hat n}_j} - 1)} .
\end{array}
\end{equation}
We find that this Hamiltonian satisfies a pseudo-Hermitian symmetry~\cite{mostafazadeh2002pseudo}
 \begin{equation}
 {{\cal K}}{\hat H_B}{{{\cal K}}^\dag } = \hat H_B^\dag,
 \end{equation}
 with ${{\cal K}} = {{{\cal R}}_z}{{\cal I}{\cal T}}$ an anti-unitary operator, 
 which guarantee that the eigenenergies must be real (as under OBCs) or come in complex-conjugate pair (as under PBCs).
 Explicitly,
 ${\cal I}$ represents the inversion symmetry that transforms $j$ to $L+1-j$, so that
 \begin{equation}
 \begin{array}{lllll}
{{H'}_B} &= {{\cal I}}{H_B}{{{\cal I}}^\dag }\\
&=  - \sum\limits_{j = 1}^{L - 1} {\left( {{J_L}\hat b_{j + 1}^\dag {e^{ - i\theta {{\hat n}_{j + 1}}}}{{\hat b}_j} + {J_R}\hat b_j^\dag {e^{i\theta {{\hat n}_{j + 1}}}}{{\hat b}_{j + 1}}} \right)}\\&+ \frac{U}{2}\sum\limits_{j = 1}^L {{{\hat n}_j}({{\hat n}_j} - 1)} . 
\end{array}
 \end{equation}
Next, ${\cal T}$ represents the time-reversal symmetry for spinless particles (i.e., a complex conjugation), which leads to
\begin{equation}
 \begin{array}{lllll}
{{H''}_B}&=  {{\cal T}}{H'}_B{{{\cal T}}^\dag } \\&=  - \sum\limits_{j = 1}^{L - 1} {\left( {{J_L}\hat b_{j + 1}^\dag {e^{i\theta {{\hat n}_{j + 1}}}}{{\hat b}_j} + {J_R}\hat b_j^\dag {e^{ - i\theta {{\hat n}_{j + 1}}}}{{\hat b}_{j + 1}}} \right)}\\&+ \frac{U}{2}\sum\limits_{j = 1}^L {{{\hat n}_j}({{\hat n}_j} - 1)} . 
\end{array}
\end{equation}
Finally, ${{{\cal R}}_z} = {e^{ - i\theta {{\hat n}_j}({{\hat n}_j} - 1)/2}}$ is a rotation operator that transforms the annihilation and creation operators as
\begin{equation}
\begin{array}{ccccc}
b_j^\dag  \to {e^{ - i\theta {{\hat n}_j}({{\hat n}_j} - 1)/2}}b_j^\dag {e^{i\theta {{\hat n}_j}({{\hat n}_j} - 1)/2}} = b_j^\dag {e^{ - i\theta {{\hat n}_j}}},\\
{b_j} \to {e^{ - i\theta {{\hat n}_j}({{\hat n}_j} - 1)/2}}{b_j}{e^{i\theta {{\hat n}_j}({{\hat n}_j} - 1)/2}} = {e^{i\theta {{\hat n}_j}}}{b_j}.
\end{array}
\end{equation}
Applying this rotation operation to the Hamiltonian, we have 

\begin{widetext}
\begin{equation}
\begin{array}{lllll}
{{H'''}_B} =  {{{\cal R}}_z}{{H''}_B}{{\cal R}}_z^\dag \\
 = - \sum\limits_{j = 1}^{L - 1} {\left( {{J_L}{{{\cal R}}_z}\hat b_{j + 1}^\dag {{\cal R}}_z^\dag {{{\cal R}}_z}{e^{i\theta {{\hat n}_{j + 1}}}}{{\hat b}_j}{{\cal R}}_z^\dag  + {J_R}{{{\cal R}}_z}\hat b_j^\dag {{\cal R}}_z^\dag {{{\cal R}}_z}{e^{ - i\theta {{\hat n}_{j + 1}}}}{{\hat b}_{j + 1}}{{\cal R}}_z^\dag } \right)+ \frac{U}{2}\sum\limits_{j = 1}^L {{{\hat n}_j}({{\hat n}_j} - 1)} } \\
 = - \sum\limits_{j = 1}^{L - 1} {\left( {{J_L}b_{j + 1}^\dag {e^{ - i\theta {{\hat n}_{j + 1}}}}{e^{i\theta {{\hat n}_{j + 1}}}}{e^{i\theta {{\hat n}_j}}}{b_j} + {J_R}b_j^\dag {e^{ - i\theta {{\hat n}_j}}}{e^{ - i\theta {{\hat n}_{j + 1}}}}{e^{i\theta {{\hat n}_{j + 1}}}}{b_{j + 1}}} \right)+ \frac{U}{2}\sum\limits_{j = 1}^L {{{\hat n}_j}({{\hat n}_j} - 1)} } \\
 = - \sum\limits_{j = 1}^{L - 1} {\left( {{J_L}b_{j + 1}^\dag {e^{i\theta {{\hat n}_j}}}{b_j} + {J_R}b_j^\dag {e^{ - i\theta {{\hat n}_j}}}{b_{j + 1}}} \right)+ \frac{U}{2}\sum\limits_{j = 1}^L {{{\hat n}_j}({{\hat n}_j} - 1)} } \\
 =H^{\dagger}_B.
\end{array}
\end{equation}
\end{widetext}
\blue{Note that the pseudo-Hermitian symmetry also ensures that the eigeneneriges must either be real or come in complex-conjugate pairs~\cite{mostafazadeh2002pseudo}.
To see this, let us assume $|\psi\rangle$ is an eigenstate of $\hat{H}_B$ with an eigenenergy $E$, 
\begin{eqnarray}
\hat{H}_B|\psi\rangle=E|\psi\rangle.
\end{eqnarray}
The pseudo-Hermitian symmetry leads to
\begin{eqnarray}
\hat{H}_B^\dagger{\cal K}|\psi\rangle={\cal K} \hat{H}_B|\psi\rangle=E{\cal K} |\psi\rangle,
\end{eqnarray}
Namely, $|\phi\rangle={\cal K} |\psi\rangle$ is an eigenstate of $\hat{H}_B^\dagger$, or in other words,
a left eigenstate of $\hat{H}_B$ with 
\begin{eqnarray}
\langle \phi| \hat{H}_B=E^*\langle \phi|.
\end{eqnarray}
Therefore, there must be a right eigenstate $|\phi'\rangle$ of $\hat{H}_B$ satisfying 
\begin{eqnarray}
\hat{H}_B|\phi'\rangle=E^*|\phi'\rangle.
\end{eqnarray}
As a conclusion, the pseudo-Hermitian symmetry ensures either a real eigenenergy $E$ if $|\psi\rangle=|\phi'\rangle$ (so that $E=E^*$), or a complex conjugate pair of $E$ and $E^*$ if $|\psi\rangle\neq|\phi'\rangle$.
}

\subsection*{Dynamical symmetry}
In the Hermitian limit of our model ($J_L=J_R$), it has been shown that \cite{Liu2018PRL} 
\begin{equation}
{\left\langle {{{\hat n}_j}\left( t \right)} \right\rangle _{ + U}} = {\left\langle {{{\hat n}_{j'}}\left( t \right)} \right\rangle _{ - U}}, \;{\left\langle {{{\hat n}_j}\left( t \right)} \right\rangle _{ + \theta }} = {\left\langle {{{\hat n}_{j'}}\left( t \right)} \right\rangle _{ - \theta }},
\end{equation}
with $j'=L+1-j$, and $\left\langle\right\rangle$ denoting the average of a Heisenberg operator on the initial state $|\Psi_0\rangle=\prod\nolimits_i {\hat b_i^\dag \left| 0 \right\rangle }$.
Yet these relations no longer hold when $J_L\neq J_R$ and the symmetry between different values of $U$ and $\theta$ need to be reexamined for the non-Hermitian Hamiltonian
\begin{equation}
\begin{array}{lllll}
{\hat H_{B}} =  & - \sum\limits_{j = 1}^{L - 1} {\left( {{J_L}\hat b_j^\dag {e^{i\theta {{\hat n}_j}}}{{\hat b}_{j + 1}} + {J_R}\hat b_{j + 1}^\dag {e^{ - i\theta {{\hat n}_j}}}{{\hat b}_j}} \right)}  \\&+\frac{U}{2}\sum\limits_{j = 1}^L {{{\hat n}_j}({{\hat n}_j} - 1)}.
\end{array}
\end{equation}
The pseudo-Hermitian symmetry operation leads to the following relations:
\begin{equation}
{{\cal K}}{e^{ - i{{\hat H}_B}t}}{{{\cal K}}^\dag } = {e^{i\hat H_B^\dag t}},\\
{{\cal K}}{n_j}{{{\cal K}}^\dag } = {{\cal I}}{n_j}{{{\cal I}}^\dag } = {\hat n_{j'}}.
\end{equation}

 Thus we obtain
 \begin{equation}
 \begin{array}{l}
{\left\langle {{{\hat n}_j}\left( t \right)} \right\rangle _{ + \theta,+U }} \equiv \left\langle {{\Psi _0}} \right|{e^{i\hat H_{B, + \theta,+U }^\dag t}}{{\hat n}_j}{e^{ - i{{\hat H}_{B, + \theta,+U }}t}}\left| {{\Psi _0}} \right\rangle \\
 = \left\langle {{\Psi _0}} \right|{{{\cal K}}^\dag }{e^{ - i{{\hat H}_{B, + \theta,+U }}t}}{{\cal K}}{{\hat n}_j}{{{\cal K}}^\dag }{e^{i\hat H_{B, + \theta,+U }^\dag t}}{{\cal K}}\left| {{\Psi _0}} \right\rangle \\
 = \left\langle {{\Psi _0}} \right|{e^{ - i{{\hat H}_{B, + \theta,+U }}t}}{{\hat n}_{j'}}{e^{i\hat H_{B, + \theta,+U }^\dag t}}\left| {{\Psi _0}} \right\rangle.
\end{array}
\end{equation}
We then consider the time-reversal operation
\begin{equation}
\begin{array}{l}
{{\cal T}}{{\hat H}_{B, + \theta,+U }}{{{\cal T}}^{ - 1}} = {{\hat H}_{B, - \theta,+U }}
\\\Rightarrow
{{\cal T}}{e^{ - i{{\hat H}_{B, + \theta , + U}}t}}{{{\cal T}}^{ - 1}} = {e^{i{{\hat H}_{B, - \theta , + U}}t}}.
\end{array}
\end{equation}
with ${\cal T}$ the complex conjugation operator for spinless particles,
and an operation
\begin{equation}
{{\cal P}}{{\hat H}_{B, + \theta , + U}}{{{\cal P}}^\dag } =  - {{\hat H}_{B, + \theta , - U}}
\Rightarrow
{{\cal P}}{e^{ - i{{\hat H}_{B, + U}}t}}{{{\cal P}}^\dag } = {e^{i{{\hat H}_{B, - U}}t}}.
\end{equation}
with ${\rm{}}\;{{\cal P}} = {e^{i\pi \sum\nolimits_r {{{\hat n}_{2r + 1}}} }}$ the number parity operator measuring the parity of total particle number on the odd sites.
Then, we have 
\begin{equation}
\begin{array}{lllll}
&{\left\langle {{{\hat n}_j}\left( t \right)} \right\rangle _{ + \theta , + U}}\\ &= \left\langle {{\Psi _0}} \right|{e^{i\hat H_{B, + \theta , + U}^\dag t}}{{\hat n}_j}{e^{ - i{{\hat H}_{B, + \theta , + U}}t}}\left| {{\Psi _0}} \right\rangle \\
 &= \left\langle {{\Psi _0}} \right|{{{\cal T}}^{ - 1}}{e^{ - i{{\hat H}^\dag }_{B, - \theta , + U}t}}{{\cal T}}{{\hat n}_j}{{{\cal T}}^{ - 1}}{e^{i{{\hat H}_{B, - \theta , + U}}t}}{{\cal T}}\left| {{\Psi _0}} \right\rangle \\
 &= \left\langle {{\Psi _0}} \right|{e^{ - i{{\hat H}^\dag }_{B, - \theta , + U}t}}{{\hat n}_j}{e^{i{{\hat H}_{B, - \theta , + U}}t}}\left| {{\Psi _0}} \right\rangle \\
 &= \left\langle {{\Psi _0}} \right|{{{\cal P}}^\dag }{e^{i{{\hat H}^\dag }_{B, - \theta , - U}t}}{{\cal P}}{{\hat n}_j}{{{\cal P}}^\dag }{e^{ - i{{\hat H}_{B, - \theta , - U}}t}}{{\cal P}}\left| {{\Psi _0}} \right\rangle \\
 &= \left\langle {{\Psi _0}} \right|{e^{i{{\hat H}^\dag }_{B, - \theta , - U}t}}{{\hat n}_j}{e^{ - i{{\hat H}_{B, - \theta , - U}}t}}\left| {{\Psi _0}} \right\rangle \\
 &= {\left\langle {{{\hat n}_j}\left( t \right)} \right\rangle _{ - \theta , - U}}.
\end{array}
\end{equation}
Thus, we have a dynamic symmetry combined the changing the sign of $\theta$ and $U$ at the same time, 
\begin{equation}
{\left\langle {{{\hat n}_j}\left( t \right)} \right\rangle _{ + \theta , + U}} =   {\left\langle {{{\hat n}_j}\left( t \right)} \right\rangle _{ - \theta , - U}}.
\end{equation}

\subsection{Perturbation explanation for asymmetric spreading}
\blue{In order to explain the asymmetric density spreading, we reproduce the derivation in Ref. \cite{Liu2018PRL} that expands the evolution operator and evaluates the interferences effect between different order. The time evolution operator can be expanded as  
\[{{\hat{\cal U}}} = {e^{ - i{{\hat H}_B}t}} = \sum\limits_{n = 0}^\infty  {\frac{{{{\left( { - i{{\hat H}_B}t} \right)}^n}}}{{n!}}}  = 1 - i{\hat H_B}t + \frac{{{{\left( {i{{\hat H}_B}t} \right)}^2}}}{{2!}} - ...\]
The initial state $\left| {{\psi _0}} \right\rangle $ is symmetric around the lattice center with $\left| {{\psi _0}} \right\rangle  = {{\cal I}}\left| {{\psi _0}} \right\rangle $ and the final state can be superimposed with product states in Fock space.  
 To show that  the interference between the different order leads to asymmetric density evolution,
 we consider a state $|\psi_1\rangle$ with an imbalanced distribution between the left and right halves of the system, and its spatial inversion 
 $\left| {{\psi _2}} \right\rangle  = {{\cal I}}\left| {{\psi _1}} \right\rangle$.
Asymmetric dynamics occurs when  $\left\langle  {{\psi }_{1}} \right|\hat{\cal U}\left| {{\psi }_{0}} \right\rangle\neq\left\langle  {{\psi }_{2}} \right|\hat{\cal U}\left| {{\psi }_{0}} \right\rangle$.
The $k$th-order expansion of these inner products can be written as 
	\[M_{k}^{\left( 1 \right)}=\left\langle  {{\psi }_{1}} \right|\frac{{{\left( -i{{{\hat{H}}}_{B}}t \right)}^{k}}}{k!}\left| {{\psi }_{0}} \right\rangle =\frac{{{\left( -it \right)}^{k}}}{k!}{{A}_{k}},\]
with ${{A}_{k}}=\left\langle  {{\psi }_{1}} \right|{{\left( {{{\hat{H}}}_{B}} \right)}^{k}}\left| {{\psi }_{0}} \right\rangle$, and
	\[M_{k}^{\left( 2 \right)}=\left\langle  {{\psi }_{2}} \right|\frac{{{\left( -i{{{\hat{H}}}_{B}}t \right)}^{k}}}{k!}\left| {{\psi }_{0}} \right\rangle =\frac{{{\left( -it \right)}^{k}}}{k!}{{B}_{k}},\]
with ${{B}_{k}}=\left\langle  {{\psi }_{2}} \right|{{\left( {{{\hat{H}}}_{B}} \right)}^{k}}\left| {{\psi }_{0}} \right\rangle$. Taking the symmetry properties into consideration, we obtain 
\begin{align}
    B_k & = \left\langle \psi_2 \right| \hat{H}_B^k \left| \psi_0 \right\rangle \notag \\
        & = \left\langle \psi_1 \right| {\cal I}^\dag \hat{H}_B^k {\cal I} \left| \psi_0 \right\rangle \notag \\
        & = e^{i(\phi_2 - \phi_0)} \left\langle \psi_1 \right| {\cal I}^\dag {\cal R}^\dag \hat{H}_B^k {\cal R} {\cal I} \left| \psi_0 \right\rangle \notag \\
        & = e^{i(\phi_2 - \phi_0)} \left\langle \psi_1 \right| {\cal T} {\cal K}^\dag \hat{H}_B^k {\cal K} {\cal T}^{-1} \left| \psi_0 \right\rangle \notag \\
        & = e^{i(\phi_2 - \phi_0)} \left( \left\langle \psi_1 \right| {\cal K}^\dag \hat{H}_B^k {\cal K} \left| \psi_0 \right\rangle \right)^*.\label{sub_Bk}
\end{align}
In the above derivation, we have used following equations:
\[
\begin{array}{l}
    {\cal K} = {\cal R} {\cal I} {\cal T} \Rightarrow {\cal K} {\cal T}^{-1} = {\cal R} {\cal I} \Rightarrow {\cal T} {\cal K}^\dag = {\cal I}^\dag {\cal R}^\dag, \\
    {\cal K} \hat{H}_B^k {\cal K}^\dag = \hat{H}_B^{k \dag} \Rightarrow \hat{H}_B^k = {\cal K}^\dag \hat{H}_B^{k \dag} {\cal K}.
\end{array}
\]
In Eq.~\eqref{sub_Bk} we have used ${\cal R}\left| {{\psi }_{0,2}} \right\rangle ={{e}^{i{{\phi }_{0,2}}}}\left| {{\psi }_{0,2}} \right\rangle .$There is difference between the Hermitian case and non-Hermitian case. The non-zero contribution of the interference between $k$th and $(k+1)$th orders 
\begin{align}
  & {{S}_{1}}=\left| M_{k}^{(1)}+M_{k+1}^{(1)} \right|=\frac{{{t}^{k}}}{k!}\left| {{A}_{k}}+\frac{-it}{k+1}{{A}_{k+1}} \right|; \notag\\ 
 & {{S}_{2}}=\left| M_{k}^{(2)}+M_{k+1}^{(2)} \right|=\frac{{{t}^{k}}}{k!}\left| {{B}_{k}}+\frac{-it}{k+1}{{B}_{k+1}} \right|,\notag \\ 
 & \Rightarrow {{S}_{2}}=\frac{{{t}^{k}}}{k!}\left| B_{k}^{*}-\frac{-it}{k+1}B_{k+1}^{*} \right| .
\end{align}
Thus, we obtain 
\begin{align}
    S_1 & = \frac{t^k}{k!} \left| \left\langle \psi_1 \right| \left( \hat{H}_B \right)^k \left| \psi_0 \right\rangle + \frac{-it}{k + 1} \left\langle \psi_1 \right| \left( \hat{H}_B \right)^{k + 1} \left| \psi_0 \right\rangle \right|,\notag \\
    S_2 & = \frac{t^k}{k!} \left| \left\langle \psi_1 \right| {\cal K}^\dag \left(\hat{H}_B\right)^k {\cal K} \left| \psi_0 \right\rangle - \frac{-it}{k + 1} \left\langle \psi_1 \right| {\cal K}^\dag  \left(\hat{H}_B\right)^{k+1} {\cal K} \left| \psi_0 \right\rangle \right|.
\end{align}
Thus,
$\left\langle  {{\psi }_{1}} \right|\hat{\cal U}\left| {{\psi }_{0}} \right\rangle\neq\left\langle  {{\psi }_{2}} \right|\hat{\cal U}\left| {{\psi }_{0}} \right\rangle$ is satisfied since ${{S}_{1}}\ne {{S}_{2}}$ in general cases, meaning that the final state is asymmetric. 
Without further restriction of the model,
the only exceptions are the Hermitian cases ($J_L=J_R$) with $\theta=0$ or $\pi$, namely for bosons or ``pseudofermions", where $S_1=S_2$ has been proven in Ref.~\cite{Liu2018PRL}.
}

\emph{{\clr Acknowledgements}.---}
This work is supported by National Natural Science Foundation of China (Grant No. 12104519 and No. 12474159) and the
Guangdong Project (Grant No. 2021QN02X073), and the China Postdoctoral Science Foundation (Grant No. 2024T171076).

\newpage

\begin{widetext}

	
\begin{center}
\textbf{\large Supplementary Materials}
\end{center}


\setcounter{equation}{0}
\renewcommand{\thesection}{Supplemental Note \arabic{section}} 
\setcounter{table}{0}   
\setcounter{figure}{0}  
\renewcommand{\thetable}{I\arabic{table}}   
\renewcommand{\thefigure}{Supplemental Figure \arabic{figure}} 
\renewcommand{\figurename}{}  

\section{Effective Hamiltonian for $N=2$ in the subspace of bound states}
For a system with $N=2$ particles, a strong Hubbard interaction ($U\gg 1$) separates two-particle bound states from others in their energies,
thus we can obtain an effective non-interacting Hamiltonian to describe the system in the subspace of bound states.
Explicitly,
we split the extended Bose-Hubbard model into unperturbed and perturbed parts, with the unperturbed Hamiltonian

\begin{eqnarray}
{\hat H_0} = \frac{U}{2}\sum\limits_{j = 1}^L {{{\hat n}_j}({{\hat n}_j} - 1)} ,
\end{eqnarray}
and the perturbation term 
\begin{eqnarray}
{H_{{\rm{hop}}}} =  - \sum\limits_{j = 1}^{L - 1} {\left( {{J_L}\hat b_j^\dag {e^{ - i\theta {{\hat n}_j}}}{{\hat b}_{j + 1}} + {J_R}\hat b_{j + 1}^\dag {e^{i\theta {{\hat n}_j}}}{{\hat b}_j}} \right)}.
\end{eqnarray}
For $N=2$ with unperturbed energy $E=U$, the eigenstates are given by
\begin{eqnarray}
\left. {\left| {{\alpha _l}} \right\rangle } \right\rangle  = \frac{1}{{\sqrt {2!} }} (\hat b_l^\dag )^2\left| {{\rm{vac}}} \right\rangle  \equiv \left| {{2_l}} \right\rangle {\rm{,   (}}i = 1,2,...,L{\rm{)}}{\rm{.}}
\end{eqnarray}
The effective Hamiltonian ${{\hat{H}}_{\text{eff}}}$ can be perturbatively obtained as  
${\hat H_{{\rm{eff}}}} = E + {{{\cal P}}_{{\rm{int}}}}{\hat H_{{\rm{hop}}}}{{{\cal P}}_{{\rm{int}}}} + {{{\cal P}}_{{\rm{int}}}}{\hat H_{{\rm{hop}}}}{\left( {E - {{\hat H}_{{\rm{int}}}}} \right)^{ - 1}}{\hat H_{{\rm{hop}}}}{{{\cal P}}_{{\rm{int}}}} + {{\cal O}}\left( {\hat H_{{\rm{hop}}}^3} \right),$ with ${{{\cal P}}_{{\rm{int}}}} = \sum\nolimits_{i = 1}^L {\left| {\left. {{\alpha _i}} \right\rangle } \right\rangle } \left\langle {\left\langle {{\alpha _i}} \right.} \right|$ the projector onto the eigenstates of ${{\hat{H}}_{\text{eff}}}$. 

The second order contribution as
\begin{eqnarray}
\left\langle \left\langle \alpha_{l'} \right| \hat{H}_{\text{hop}} (E - \hat{H}_{\text{int}})^{-1} \hat{H}_{\text{hop}} \left| \alpha_l \right\rangle \right\rangle
\end{eqnarray}
%
%
The first perturbation term should be zero, since no term makes the paired particles to move together. The second order correction term is 
\begin{eqnarray}
\left\langle {\left\langle {{\alpha _{l'}}} \right.} \right|{\hat H_{{\rm{hop}}}}{\left( {E - {{\hat H}_{{\rm{int}}}}} \right)^{ - 1}}{\hat H_{{\rm{hop}}}}\left| {\left. {{\alpha _l}} \right\rangle } \right\rangle  = \left\langle {{\rm{vac}}} \right|\frac{1}{{\sqrt {2!} }}{\hat b_{l'}}^2{\hat H_{{\rm{hop}}}}{\left( {E - {{\hat H}_{{\rm{int}}}}} \right)^{ - 1}}{\hat H_{{\rm{hop}}}}\frac{1}{{\sqrt {2!} }}\hat b{_l^\dag{}^2}\left| {{\rm{vac}}} \right\rangle .
\end{eqnarray}
A straightforward calculation leads to 

\begin{align}
&{{\hat H}_{{\rm{hop}}}}{\left( {E - {{\hat H}_{{\rm{int}}}}} \right)^{ - 1}}{{\hat H}_{{\rm{hop}}}}\nonumber\\
& = {{\hat H}_{{\rm{hop}}}}{\left( {E - \frac{U}{2}\sum\limits_{j = 1}^L {{{\hat n}_j}({{\hat n}_j} - 1)} } \right)^{ - 1}}{{\hat H}_{{\rm{hop}}}}\nonumber\\
 &= \sum\limits_{j = 1}^{L - 1} {\left( {{J_L}\hat b_j^\dag {e^{ - i\theta {{\hat n}_j}}}{{\hat b}_{j + 1}} + {J_R}\hat b_{j + 1}^\dag {e^{i\theta {{\hat n}_j}}}{{\hat b}_j}} \right)} {\left( {E - \frac{U}{2}\sum\limits_{j' = 1}^L {{{\hat n}_{j'}}({{\hat n}_{j'}} - 1)} } \right)^{ - 1}}\sum\limits_{j'' = 1}^{L - 1} {\left( {{J_L}\hat b_{j''}^\dag {e^{ - i\theta {{\hat n}_{j''}}}}{{\hat b}_{j'' + 1}} + {J_R}\hat b_{j'' + 1}^\dag {e^{i\theta {{\hat n}_{j''}}}}{{\hat b}_{j''}}} \right)} .
\end{align}

Calculating step by step, we obtain  
\begin{align}
&\sum\limits_{j'' = 1}^{L - 1} {\left( {{J_L}\hat b_{j''}^\dag {e^{ - i\theta {{\hat n}_{j''}}}}{{\hat b}_{j'' + 1}} + {J_R}\hat b_{j'' + 1}^\dag {e^{i\theta {{\hat n}_{j''}}}}{{\hat b}_{j''}}} \right)} \frac{1}{{\sqrt {2!} }}\hat b{_l^\dag {}^2}\left| {{\rm{vac}}} \right\rangle {\rm{  }}\left[ {\hat b\left| n \right\rangle  = \sqrt n \left| {n - 1} \right\rangle ,{{\hat b}^\dag }\left| n \right\rangle  = \sqrt {n + 1} \left| {n + 1} \right\rangle } \right]\nonumber\\
 &= \sqrt 2 {J_L}\left| {{1_{l - 1}}{1_l}} \right\rangle  + \sqrt 2 {e^{i\theta }}{J_R}\left| {{1_l}{1_{l + 1}}} \right\rangle ,
\end{align}

and 

\begin{align}
&{\left( {E - \frac{U}{2}\sum\limits_{j' = 1}^L {{{\hat n}_{j'}}({{\hat n}_{j'}} - 1)} } \right)^{ - 1}}\sum\limits_{j'' = 1}^{L - 1} {\left( {{J_L}\hat b_{j''}^\dag {e^{ - i\theta {{\hat n}_{j''}}}}{{\hat b}_{j'' + 1}} + {J_R}\hat b_{j'' + 1}^\dag {e^{i\theta {{\hat n}_{j''}}}}{{\hat b}_{j''}}} \right)}\frac{1}{{\sqrt {2!} }}\hat b{_l^\dag {}^2}\left| {{\rm{vac}}} \right\rangle  \nonumber \\
 &= \frac{1}{U}\left[ {\sqrt 2 {J_L}\left| {{1_{l - 1}}{1_l}} \right\rangle  + \sqrt 2 {e^{i\theta }}{J_R}\left| {{1_l}{1_{l + 1}}} \right\rangle } \right].
\end{align}

Further calculation leads to 
\begin{align}
&\sum\limits_{j = 1}^{L - 1} {\left( {{J_L}\hat b_j^\dag {e^{ - i\theta {{\hat n}_j}}}{{\hat b}_{j + 1}} + {J_R}\hat b_{j + 1}^\dag {e^{i\theta {{\hat n}_j}}}{{\hat b}_j}} \right)} {\left( {E - \frac{U}{2}\sum\limits_{j' = 1}^L {{{\hat n}_{j'}}({{\hat n}_{j'}} - 1)} } \right)^{ - 1}}\sum\limits_{j'' = 1}^{L - 1} {\left( {{J_L}\hat b_{j''}^\dag {e^{ - i\theta {{\hat n}_{j''}}}}{{\hat b}_{j'' + 1}} + {J_R}\hat b_{j'' + 1}^\dag {e^{i\theta {{\hat n}_{j''}}}}{{\hat b}_{j''}}} \right)} \nonumber \\
&\times \frac{1}{{\sqrt {2!} }}\hat b{_l^\dag {}^2}\left| {{\rm{vac}}} \right\rangle   \\
& = \sum\limits_{j = 1}^{L - 1} {\left( {{J_L}\hat b_j^\dag {e^{ - i\theta {{\hat n}_j}}}{{\hat b}_{j + 1}} + {J_R}\hat b_{j + 1}^\dag {e^{i\theta {{\hat n}_j}}}{{\hat b}_j}} \right)} \frac{1}{U}\left[ {\sqrt 2 {J_L}\left| {{1_{l - 1}}{1_l}} \right\rangle  + \sqrt 2 {e^{i\theta }}{J_R}\left| {{1_l}{1_{l + 1}}} \right\rangle } \right]\nonumber\\
& = \frac{1}{U}\sqrt 2 J_L^2\left| {{1_{l - 2}}{1_l}} \right\rangle  + \frac{1}{U}2J_L^2{e^{ - i\theta }}\left| {{2_{l - 1}}} \right\rangle  + \frac{1}{U}\sqrt 2 {e^{i\theta }}{J_L}{J_R}\left| {{1_{l - 1}}{1_{l + 1}}} \right\rangle  + \frac{1}{U}2{e^{i\theta }}{e^{ - i\theta }}{J_L}{J_R}\left| {{2_l}} \right\rangle \nonumber\\
& + \frac{1}{U}\sqrt 2 {J_L}{J_R}\left| {{1_{l - 1}}{1_{l + 1}}} \right\rangle  + \frac{1}{U}2{J_L}{J_R}\left| {{2_l}} \right\rangle  + \frac{1}{U}\sqrt 2 {e^{i\theta }}{J_R}{J_R}\left| {{1_l}{1_{l + 2}}} \right\rangle  + \frac{1}{U}2{e^{i\theta }}{J_R}{J_R}\left| {{2_{l + 1}}} \right\rangle .
\end{align}

Thus, the matrix elements can be written as 
\begin{eqnarray}
\left\langle {\left\langle {{\alpha _{l'}}} \right.} \right|{\hat H_{{\rm{hop}}}}{\left( {E - {{\hat H}_{{\rm{int}}}}} \right)^{ - 1}}{\hat H_{{\rm{hop}}}}\left| {\left. {{\alpha _l}} \right\rangle } \right\rangle  = \frac{{4{J_L}{J_R}}}{U}{\delta _{ll'}} + \frac{1}{U}2J_L^2{e^{ - i\theta }}{\delta _{l',l - 1}} + \frac{1}{U}2{e^{i\theta }}{J_R}{J_R}{\delta _{l',l + 1}}.
\end{eqnarray}

Finally, we arrive at this effective Hamiltonian, 

\begin{align}
{{\hat H}_{{\rm{eff}}}} & =U+ \sum\limits_l {\left[ {\frac{{4{J_L}{J_R}}}{U}\left| {\left. {{\alpha _l}} \right\rangle } \right\rangle \left\langle {\left\langle {{\alpha _l}} \right.} \right| + \frac{1}{U}2J_L^2{e^{ - i\theta }}\left| {\left. {{\alpha _{l - 1}}} \right\rangle } \right\rangle \left\langle {\left\langle {{\alpha _l}} \right.} \right| + \frac{1}{U}2{e^{i\theta }}{J^2_R}\left| {\left. {{\alpha _{l + 1}}} \right\rangle } \right\rangle \left\langle {\left\langle {{\alpha _l}} \right.} \right|} \right]}.
\end{align}

Since ${\hat H}_{{\rm{eff}}}$ effectively describes a non-interacting Hamiltonian in the subspace of $|\alpha\rangle\rangle$, we can apply the Fourier transformation and obtain
\begin{eqnarray}
\hat{H}_{\rm eff}(k)=U+\frac{4J_LJ_R}{U}+\frac{2J_L^2}{U}e^{i(k-\theta)}+\frac{2J_R^2}{U}e^{-i(k-\theta)}.
\end{eqnarray}
We can see from $\hat{H}_{\rm eff}(k)$ that its eigenenergies form an ellipse in the complex energy plane [as in Fig. 1(b1) and (d1) in the main text], with the anyonic statistical angle $\theta$ changing only the center of the ellipse (and modifying the quasi-momentum $k$).

In addition, the above observation also suggests that the NHSE of bound states depends only on the ratio of 
\begin{eqnarray}
\left| {2J_L^2{e^{ - i\theta }}} \right|/ {\left| {2{e^{i\theta }}J_R^2} \right|}= J_L^2/J_R^2,
\end{eqnarray}
and is unaffected by the anyonic statistic.
~In \ref{fig:Bound}, we plot the density distributions for each eigenstate at $\theta=0$ (blue line) and $\theta=-\pi/2$ (red line), which indeed are seen to be identical. 

\begin{figure}[htbp]
\centering
\includegraphics[width=0.7\linewidth]{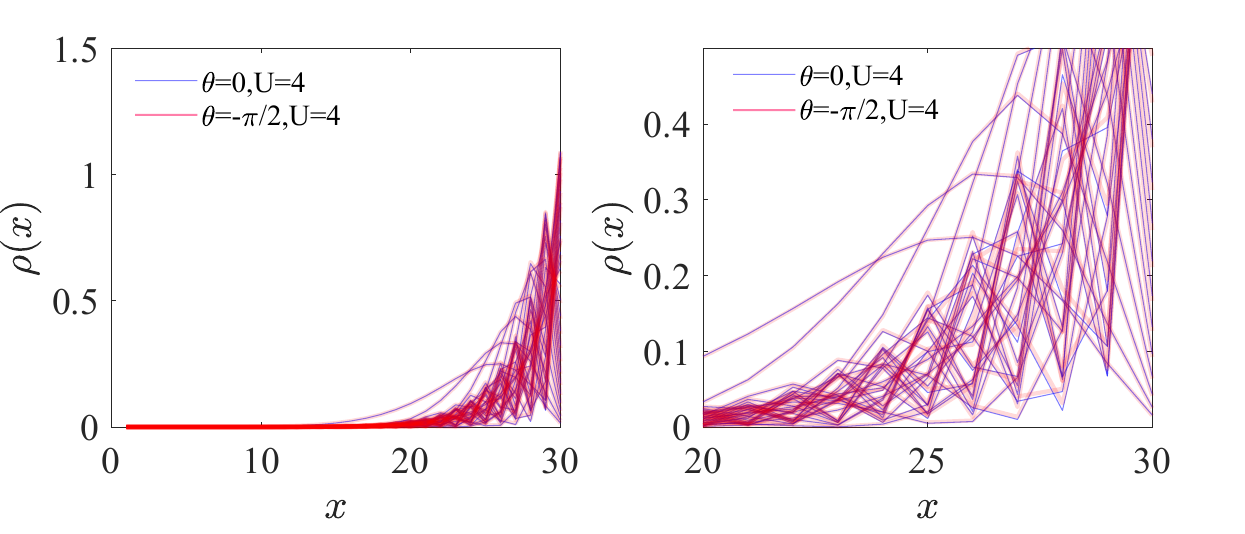}
\caption{\label{fig:Bound} 
Density distributions for $U=4$ in bound state subspace with $\theta=0$ (blue line) and $\theta=-\pi/2$ (red line). The right panel is a zoom-in of the left panel. The density distributions for two cases are almost overlapped. $\alpha=0.1$, $N=2$, and $L=30$.
}
\end{figure}   

\section{ Particle transportation}

\subsection{ $N=2$ particles for $\theta=\pi/2$}
In Fig. 2(c) of the main text, 
reversed density pumping toward the right is also observed (indicated by nonzero $\Delta t$) when $\theta \in (0, \pi)$, where the asymmetric anyonic transport shares the same left-moving tendency as the NHSE~\cite{Liu2018PRLSM}.
In \ref{fig:ChangeU}(a) to (d),
we consider two anyons with $\theta=\pi/2$ as the initial state, 
where reversed pumping is barely seen at the beginning of the evolution with different values of $U$.
However, due to the interference between the two particles, their wavefunction is seen to split into different ``branches".
Consequently, small fluctuation is seen for $\Delta N$ whenever a branch crosses the center of the lattice.
With $U$ increased, the fluctuation is seen to become stronger due to the interaction-induced diffusive effect~\cite{Ronzheimer2013PRLSM}, resulting in a decreasing of $\Delta N$ for a short period of time [e.g., $U=4,6,8$ in \ref{fig:ChangeU}(e)].
Such a reversed pumping is seen occurs when $t\gtrsim 1$, corresponding to the left-moving tendency of a minor branch (indicated by white arrows in the figure).
In contrast, as shown in Fig. 2 of the main text, the one induced by anyonic asymmetric transport with $\theta\in[-\pi,0]$ occurs at $t\lesssim1$, as it is resulted from the reversed pumping of a major branch [the left-most peak in Fig. 2(a3) to (d3) in the main text].

\begin{figure}[htbp]
\centering
\includegraphics[width=0.7\linewidth]{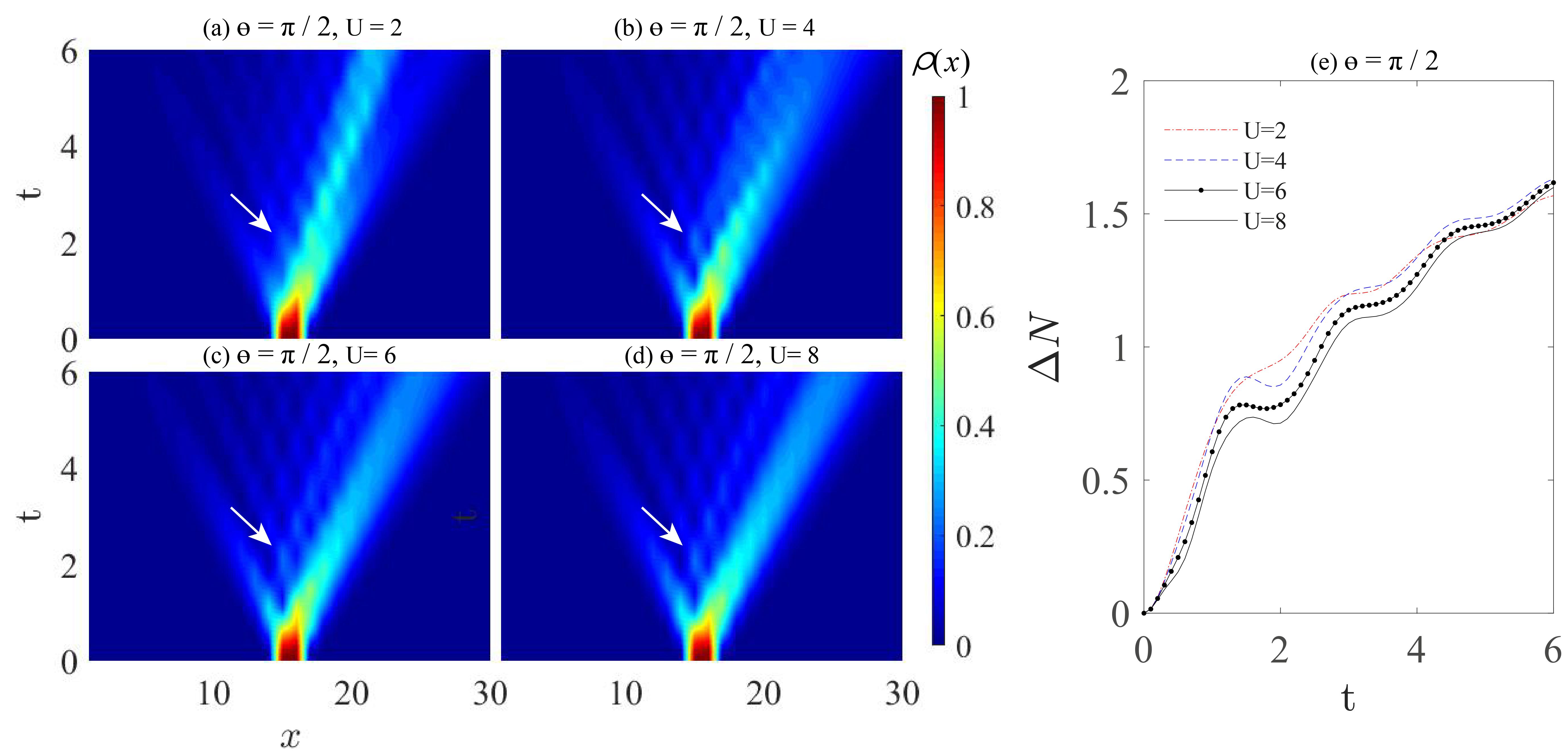}
\caption{\label{fig:ChangeU} 
Density evolutions and $\Delta N$ for $\theta=\pi/2$, with (a) $U=2$, (b) $U=4$, (c) $U=6$, and (d) $U=8$. 
(e) $\Delta N$ versus $t$ for $U=2,4,6,8$, indicated by red dotted line, blue dashed line, black dot-line, and black line, respectively. White arrows in (a) to (d) indicate the minor left-moving branches which cause the fluctuation (reversed pumping for $U=4,6,8$) of $\Delta N$ during $1\lesssim t\lesssim2$ in (e).
$\alpha=0.1$, $N=2$, and $L=30$.
}
\end{figure}   

\subsection{$N=2$ particles with different arrangements}\label{sec:diff_arrange}
In Fig. 1 of the main text, we display the evolution of two particles initially arranged next to each other, and the reversed dynamical pumping is most clearly seen at $\theta=-\pi/2$ and $U\approx 4$, as replotted in  \ref{fig_DifferentStates}(a).
Originated from the asymmetric anyonic dynamics, the reversed pumping may be less significant, or even disappear, 
if the particles are initially separated from each other by at least one site [thus acting as single particles, see Fig. \ref{fig_DifferentStates}(b)], 
located at the same lattice site [thus acting as bosons with $\theta$ = 0, \ref{fig_DifferentStates}(c)].
In \ref{fig_DifferentStates}(d) we further demonstrate the
time-dependent density imbalance $\Delta N$ for different initial states,
where the reversed pumping process can be clearly seen only for the initial state of $|a_{15} a_{16}\rangle$.
\begin{figure}[htbp]
\centering
\includegraphics[width=0.9\linewidth]{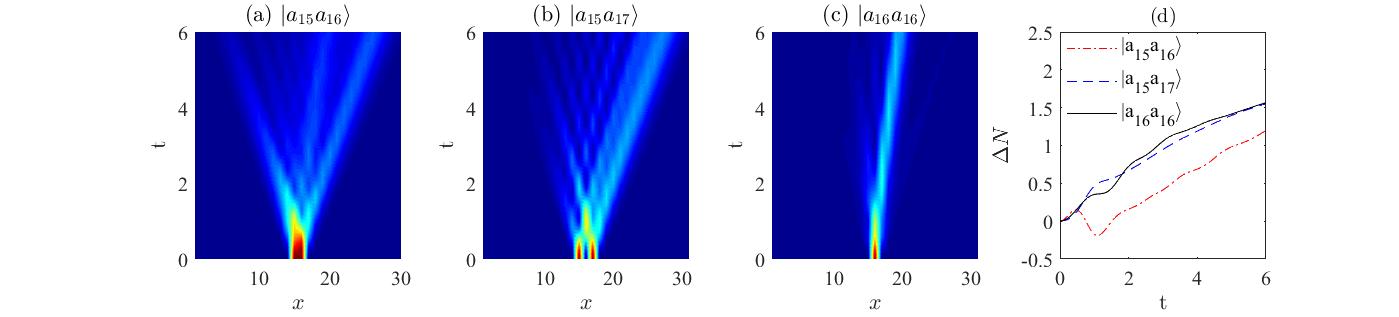}
\caption{\label{fig_DifferentStates}
The density evolution for initial states with$N=2$ particles initially located (a) next to each other, (b) separated from each other, and (c) at the same lattice site.
(d) The time-dependent density imbalance $\Delta N$ for the initial states in (a) to (c). Other parameters are $\theta=-\pi/2$ and $U=4$.
}
\end{figure} 


\subsection{Reversed density pumping under weak non-Hermicity}
\blue{In Fig. 2(a) in the main text, we show that a reversed pumping emerges even under a relatively strong non-Hermitian pumping strength with $\alpha=0.3$. 
For a much smaller $\alpha$, the pumping of NHSE will be fully suppressed during a short period of time. As can be seen in Fig.~\ref{fig:SmallAlpha}(a),
the density imbalance $\Delta N$ with $\alpha=0.001$ does not deviate much from the Hermitian case with $\alpha=0$, both of which always have negative $\Delta N$ in the time period we consider. In contrast, for $\alpha=0.01$, $\Delta N$ increases rapidly and becomes positive after $t\geqslant10$, indicating the dominance of NHSE.
In contrast, no suppression of NHSE is observed for bosons, as shown in Fig.~\ref{fig:SmallAlpha}(b).}
\begin{figure}[!htbp]
	\begin{center}		\includegraphics[width=1\linewidth]{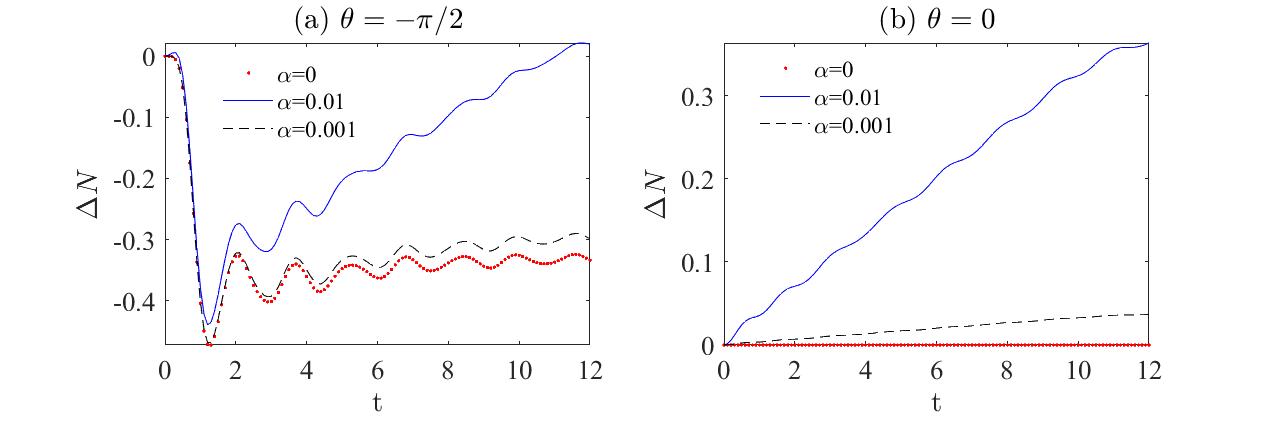}
 		\par\end{center}
 	\protect\caption{\label{fig:SmallAlpha}
Evolution of density imbalance $\Delta N$  for $N=2,L=40,U=4$, (a) $\theta=-\pi/2$ and (b) $\theta=0$, with $\alpha=0, 0.01,0.001$. 
The curve of $\alpha=0.001$ does not deviate much from the one with $\alpha=0$ in (a), indicating a domination of anyonic dynamics over the NHSE during a short-time evolution.
}
\end{figure}

\subsection{Particle transportation of $N=3$ and $N=4$}
In \ref{fig:Dynamics}, we give the time evolution for $N=3$ and $N=4$ particles with different statistical angle $\theta$ and interaction strength $U$. 
It can be seen that the tendency for density evolution is the same for the case of $N=2$ in the main text. 
Note that the seemingly ballistic evolution with a smaller velocity for $N=2$ with $\theta=-\pi/2$ and $U=0$ [Fig. 1(c3) in the main text]
cannot be observed when the particle number increases to  $N=3$ and $4$ [\ref{fig:Dynamics}(c) and (g)].
\begin{figure*}[htbp]
\centering
\includegraphics[width=0.9\linewidth]{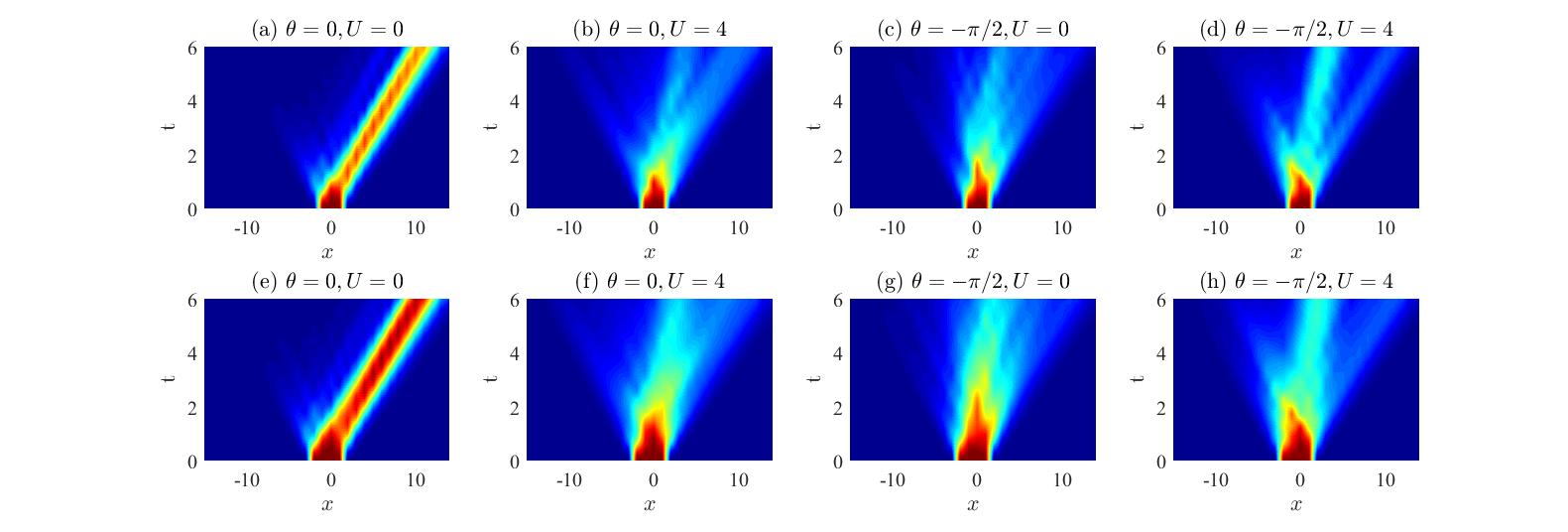}
\caption{\label{fig:Dynamics}Density evolutions for (a)-(d) $N=3$ and (e)-(h) $N=4$ particles evenly distributed at the center of the chain.
The statistical angles $\theta$ and onsite interaction $U$ are marked in each panel. 
Other parameters are $J_L=e^{-\alpha},J_R=e^{\alpha}, \alpha=0.1$ and $L=30$.}
\end{figure*}  

\subsection{Longer time evolution for various $N$}

In Fig. 3 of the main text, we display the density imbalance $\Delta N$ as a function of time $t$, where the slope for $N=2$ particles is seen to be smaller than that of the others ($N=3,4,6$). We infer that it is because for the initial state we consider with larger $N$, more particles distribute on the left half of the lattice, and  farther from the lattice's center.
Consequently, it takes longer time for the many-body wave-functions to be mostly pumped to the right half of the lattice, after which the slope of $\Delta N$ decreases.
This can be seen from \ref{fig:N2346_Long}, where we show the numerical results of $\Delta N$ as a function of $t$, with the same parameters as in Fig. 3(b) in the main text, but longer evolution time.

\begin{figure}[htbp]
\centering
\includegraphics[width=0.3\linewidth]{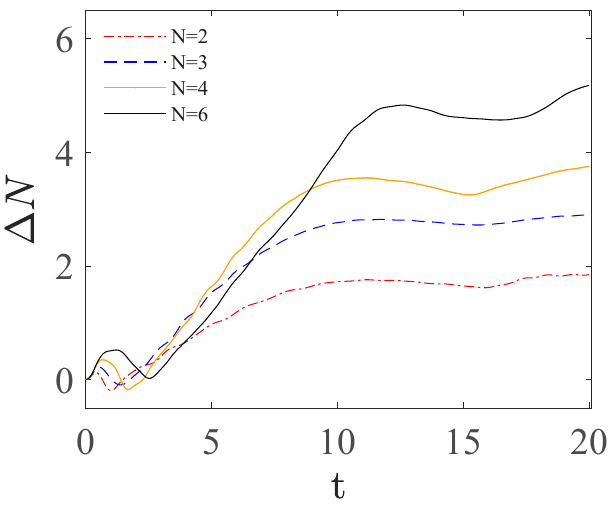}
\caption{\label{fig:N2346_Long} 
Density imbalance $\Delta N$ as a function of time $t$ for different particle numbers $N$, with $\theta=-\pi/2$ and $\alpha=0.1$.
The system's size is chosen to be $L=30$ for $N=2,4,6$, and $L=31$ for $N=3$. 
In the latter case, the density at the center of the system ($j=16$) is excluded when calculating $\Delta N$.
Namely, all the parameters and setting are chosen to be the same as in Fig. 3(b) in the main text, which only display the results up to $t=6$ for a clearer demonstration of the reversed pumping process.
}
\end{figure}   

\section{Topological invariant}
In the main text, we state that the non-Hermitian skin effect (NHSE) for our non-Hermitian anyon-Hubbard model can be characterized by a many-body spectral winding number due to a $U(1)$ symmetry, i.e. $\left[ {\hat H_B,\hat N} \right] = 0$ with $\hat{N}$ the particle number operator~\cite{Kawabata2022PRBSM}. In this section, we provide some examples to demonstrate the correspondence between NHSE and the winding number. 
Explicitly, we introduce a phase factor $\phi$ to the hopping amplitudes connecting the two ends of the 1D model ($J_{L/R}\rightarrow e^{i\phi}J_{L/R}$), and the eigenenergies form loops in the complex plane when $\phi$ changes from $0$ to $2\pi$.
The winding number is thus defined as
\begin{equation}
W(E_b):=\oint_{0}^{2\pi }{\frac{d\phi }{2\pi i }}\frac{d}{d\phi }\log \det \left[ \hat{H}_B(\phi )-E_b \right],
\end{equation}
where $E_b$ is a reference energy.
As shown in \ref{fig_Toponum}, if the reference energy point locates inside (out of) the circle of the complex energy spectrum by varying $\phi$ form $0$ to $2\pi$, the winding number $W$ will take an non-zero integer (zero), which can be read out from the evolution of argument arg[det$(\hat{H}_B(\phi)-E_b)$] versus $\phi$.

\begin{figure}[htbp]
\centering
\includegraphics[width=0.73\linewidth]{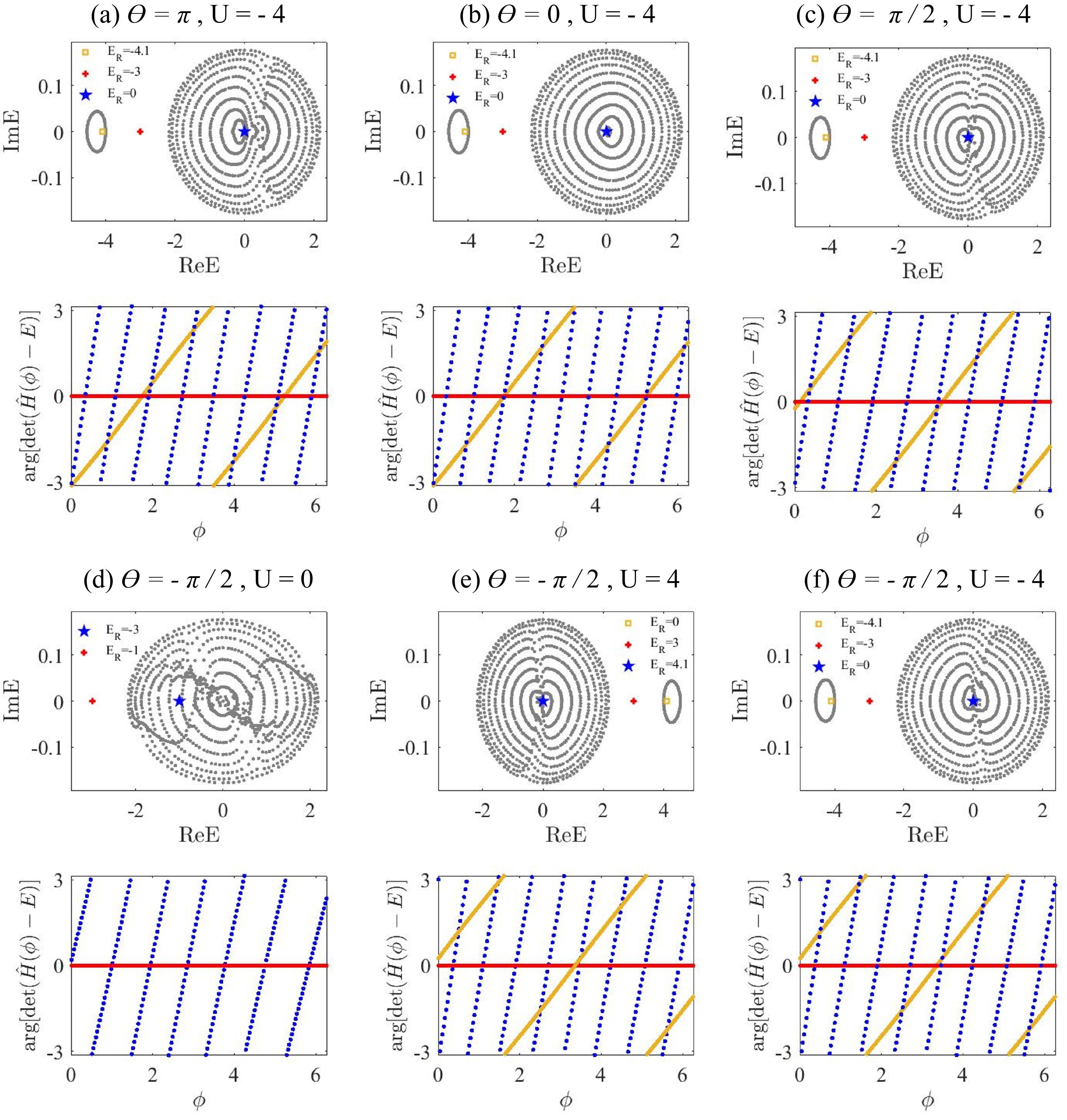}
\caption{\label{fig_Toponum} The energy spectra and arguments arg[det$(\hat{H}_B(\phi)-E_b)$] for various values of $\theta$ and $U$, as marked in each panel.  The energy spectrum is calculated under periodic boundary conditions with $\phi \in(0,2\pi/2,4\pi/5,8\pi/5)$. 
In each case, we consider three reference energies with different spectral winding numbers, marked by red square, blue pus, and black stars, respectively.
The spectral winding number $W$
can be directly read out from argument's evolution with the phase factor $\phi$ of the $U(1)$ gauge field.
Namely, we have
(a) $W=8,0,2$ with $E_b=0,-3,-4.1$,
(b) $W=9,0,2$ with $E_b=0,-3,-4.1$, 
(c) $W=9,0,2$ with $E_b=0,-3,-4.1$,
(d) $E_b=-1,-3$ with $W=6,0$,
(e) $E_b=0,3,4.1$ with $W=8,0,2$,
(f) $E_b=0,-3,-4.1$ with $W=8,0,2$,
for blue, red, and yellow colors respectively.
Other parameters $J_L=1,J_R=1.2$ and $L=20,N=2$. }
\end{figure}

\section{Two-particle correlation for $N=3$}
In the main text, we provide numerical results of the density-density correlation for $N=2$ particles to 
demonstrate the different diffusive and unidirectional dynamics in the system.
The correlation is denoted as $\Gamma_{qr}$ with $q$ and $r$ denoting the positions of two lattice sites.
The same quantity for $N=3$ is shown in \ref{fig:N3TwoP}, which exhibit similar properties as Fig. 2(d) in the main text.
That is,
the diagonal correlation $\Gamma_{qq}$ moves only toward $q=L$ when $U=0$, but shows an opposite tendency toward $q=1$ with nonzero $U$, reflecting the reversed density pumping for the latter case.
The diffusive dynamics reflected by nonzero off-diagonal correlation, $\Gamma_{qr}$ with $q\neq r$, also becomes more significant with larger $U$.

%

\begin{figure}[htbp]
\centering
\includegraphics[width=0.7\linewidth]{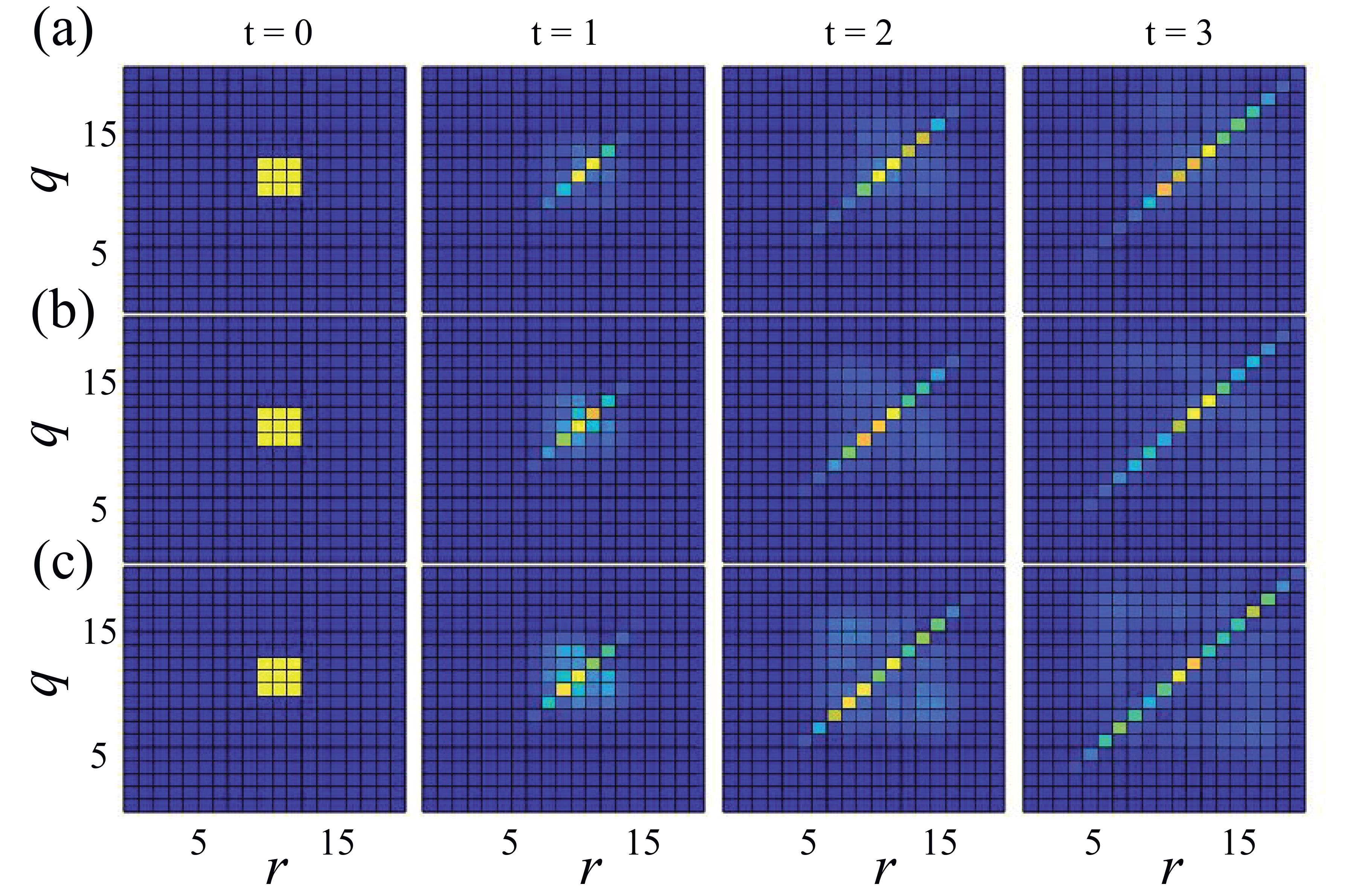}
\caption{\label{fig:N3TwoP} 
Density-density correlation $\Gamma_{qr}$ at different evolving time $t$, with $\theta=-\pi/2$, $\alpha=0.1$, $N=3$, and $L=20$.
}
\end{figure}

\section{Information spreading}
\subsection{Definition of OTOC}
In this section, we 
we derive the anyonic OTOC in the main text. 

The OTOC of anyons is defined as 
\begin{eqnarray}
{C_{jk}}(t) = {\left\langle {{{\left| {{{\left[ {{{\hat a}_j}(t),{{\hat a}_k}(0)} \right]}_\theta }} \right|}^2}} \right\rangle _\beta },
\end{eqnarray}

where $\beta$ is the inverse temperature and ${\left\langle {\hat{O}} \right\rangle _\beta }$ means the thermal ensemble average 
$\left(e^{ - \beta {{\hat H}_A}}\hat{O}\right)/{\rm{Tr}}\left( {{e^{ - \beta {{\hat H}_A}}}} \right)$ of an operator $\hat{O}$. 
Thus ${C_{jk}}(0)=0$ and increases as quantum information spreads from $k$ to $j$ site~\cite{Shen2017PRBSM,Luitz2017PRBSM,KeyserlingkPRX2018SM}. 
A straightforward calculation leads to 
\begin{eqnarray}
&&{\left| {{{\left[ {{{\hat a}_j}(t),{{\hat a}_k}(0)} \right]}_\theta }} \right|^2}\nonumber\\
&&=  {\left| {{{\hat a}_j}(t){{\hat a}_k}(0) - {e^{ - i\theta {\mathop{\rm sgn}} (j - k)}}{{\hat a}_k}(0){{\hat a}_j}(t)} \right|^2}\nonumber\\
&&=   {\left( {{{\hat a}_j}(t){{\hat a}_k}(0) - {e^{ - i\theta {\mathop{\rm sgn}} (j - k)}}{{\hat a}_k}(0){{\hat a}_j}(t)} \right)^\dag }\left( {{{\hat a}_j}(t){{\hat a}_k}(0) - {e^{ - i\theta {\mathop{\rm sgn}} (j - k)}}{{\hat a}_k}(0){{\hat a}_j}(t)} \right)\nonumber\\
&&=   \left( {\hat a_k^\dag (0)\hat a_j^\dag (t) - {e^{i\theta {\mathop{\rm sgn}} (j - k)}}\hat a_j^\dag (t)\hat a_k^\dag (0)} \right)\left( {{{\hat a}_j}(t){{\hat a}_k}(0) - {e^{ - i\theta {\mathop{\rm sgn}} (j - k)}}{{\hat a}_k}(0){{\hat a}_j}(t)} \right)\nonumber\\
&&=   \hat a_k^\dag (0)\hat a_j^\dag (t){{\hat a}_j}(t){{\hat a}_k}(0) - {e^{ - i\theta {\mathop{\rm sgn}} (j - k)}}\hat a_k^\dag (0)\hat a_j^\dag (t){{\hat a}_k}(0){{\hat a}_j}(t){ - ^{i\theta {\mathop{\rm sgn}} (j - k)}}\hat a_j^\dag (t)\hat a_k^\dag (0){{\hat a}_j}(t){{\hat a}_k}(0)\nonumber\\
&&+ \hat a_j^\dag (t)\hat a_k^\dag (0){{\hat a}_k}(0){{\hat a}_j}(t).
 \end{eqnarray}

In the main text, we have shown numerical results of the out-of-time-ordered part of the commutator based on the anyonic model in Fig. 4, given by
\begin{eqnarray}
{{\bar F}^A_{jk}}(t) =   \left\langle \hat a_j^\dag (t)\hat a_k^\dag (0){{\hat a}_j}(t) {{{\hat a}_k}(0)} \right\rangle _{A,\beta }e^{{i\theta {\mathop{\rm sgn}} (j - k)}},
\end{eqnarray}
where the subscript $A$ denotes that the ensemble average is calculated based on the anyonic Hamiltonian $\hat{H}_A$.

\subsection{OTOC at $\theta=0$}
\blue{To further demonstrate the competition between anyonic and non-Hermitian asymmetric dynamics, 
here we numerically calculate the OTOC for a non-Hermitian Bose-Hubbard model by setting $\theta =0$, where the former is absent. 
The results in Fig.~\ref{fig:Boson} and Fig.~\ref{fig:BosonAnyonsmall} show that for bosons the ensemble OTOC tends to be right side even with a very weak non-Hermicity.
In Fig.~\ref{fig:BosonAnyonsmall}, we compare the cases with $\theta=\pi/2$ (anyons) and $\theta=0$ (bosons) in the presence of a weak non-Hermiticity,
whose OTOC shows a trend to the left and right, respectively.
This is because the OTOC spreading of bosons is symmetric in the Hermitian case, and cannot suppress NHSE.
As for a single state, it is seen in Fig.~\ref{fig:Boson}(d) that the spreading of OTOC is always symmetric, reflecting the bosonic characteristics unaffected by NHSE.}


\begin{figure}[!htbp]
	\begin{center}
		\includegraphics[width=1\linewidth]{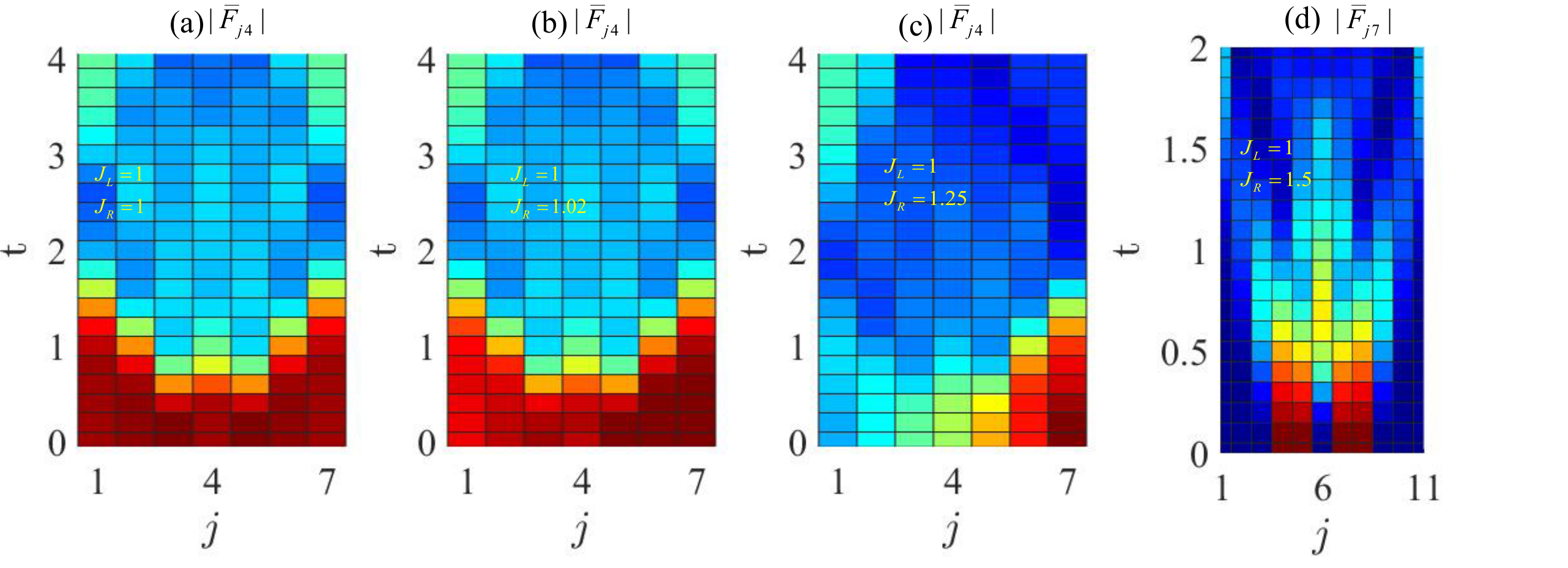}

 		\par\end{center}
 	\protect\caption{\label{fig:Boson}
 Information spreading for $\theta=0,U=4$. (a) to (c) OTOC for a thermal ensemble at finite temperature with (a) $J_R=1.0$, (b)$J_R=1.02$, and (c) $J_R=1.25$. The system has $L=7$ lattice sites and  $N=4$ particles, with the inverse temperature $\beta=1/6$. The information spreading tends to move towards the right when $J_R>J_L$. 
(d) OTOC for a single state with $L=11$, $U=4$, and $J_R=1.5$. 
The initial state is chosen as $\psi(0)\rangle = |\hat{a}_4\hat{a}_5\hat{a}_6\hat{a}_7\hat{a}_8\rangle$. Results are normalized by setting $\max_{j,t}|\bar F_{j,4}|=1$ in (a) to (c), 
$\max_{t}|F_{j,6}|=1$ in (d).}
\end{figure}
\begin{figure}[htbp]
	\begin{center}				\includegraphics[width=0.5\linewidth]{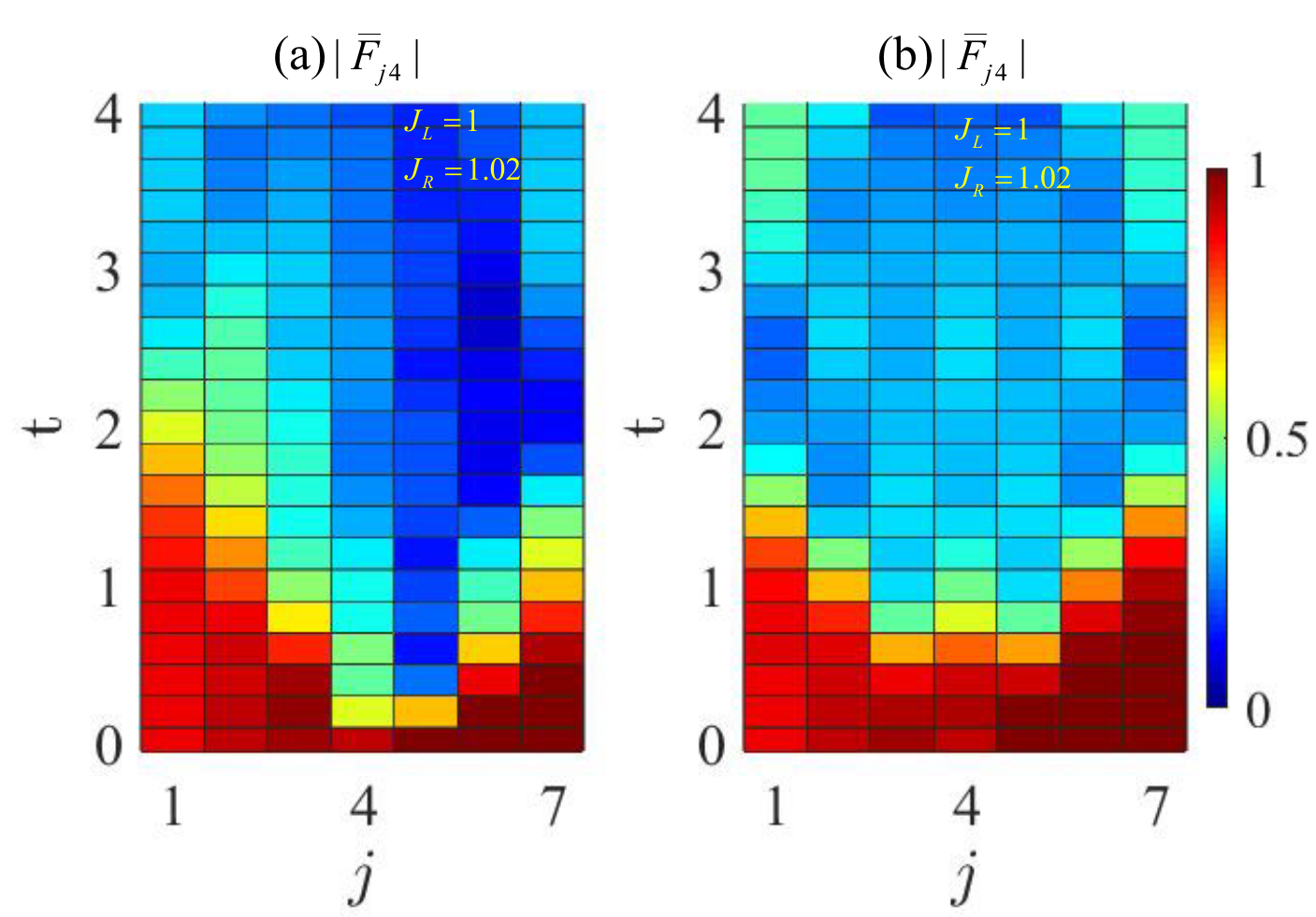}
 		\par\end{center}
 	\protect\caption{\label{fig:BosonAnyonsmall}
OTOC for a thermal ensemble at finite temperature with (a) $J_R=1.02,~\theta=0$, (b)$J_R=1.02,~\theta=-\pi/2$, . The system has $L=7$ lattice sites and  $N=4$ particles, with the inverse temperature $\beta=1/6$ and $U=4$. 
}
\end{figure}


\section{Effect of large interaction on the reversed pumping}
In the main text, we have considered the interaction strength up to $U=15$. It should be noted that further increasing the onsite interaction $U$ tends to suppress the reversed pumping, which is confirmed by our numerical calculations, especially for smaller particle number $N$, as shown in \ref{fig:FigN2346}. 

\begin{figure}[htbp!]
\centering
\includegraphics[width=0.48\linewidth]{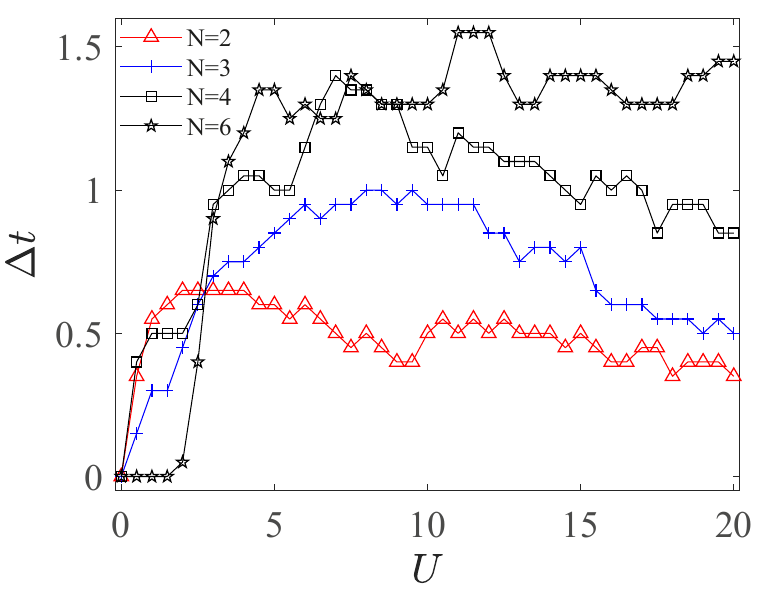}
\caption{\label{fig:FigN2346}  
Reversed pumping time $\Delta t$ for different particle numbers $N=2,3,4,6$, which are marked by different symbols and colors, as indicated in the figure.
Other parameters are $\theta=-\pi/2$ and $\alpha=0.1$.  
The system's size is chosen to be $L=30$ for $N=2,4,6$, and $L=31$ for $N=3$. In the latter case, the density at the center of the system ($j=16$) is excluded when determining $\Delta t$.}  
\end{figure}

\section{State dynamics in the hardcore limit}
\blue{In the hardcore limit of our model with only nearest neighbor hopping, state dynamics is not affected by the anyonic statistic angle, as a particle cannot hop over another to acquire the statistical phase.
As can be seen in Fig.~\ref{fig:Hardcore}, in the hardcore limit, 
despite diffusion of particles is seen in the earlier stage of the evolution (as for normal bosons),
no reversed pumping is observed even at $\theta=-\pi/2$, where the reversed pumping is the most significant for softcore anyons.
}
\begin{figure}[!htbp]
	\begin{center}
		\includegraphics[width=.45\linewidth]{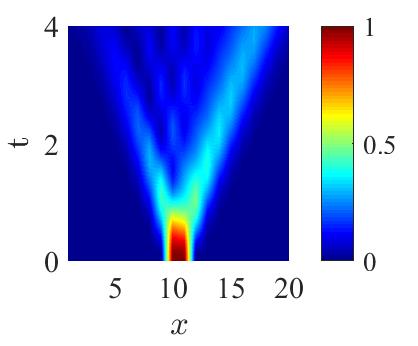}
				\includegraphics[width=.45\linewidth]{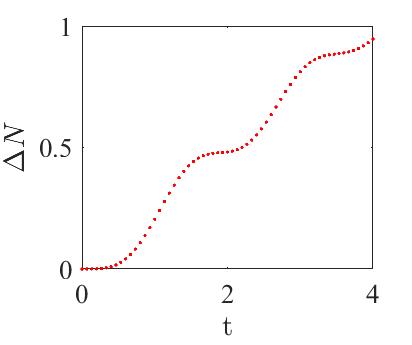}
 		\par\end{center}
 	\protect\caption{\label{fig:Hardcore}
Time evolution in the hardcore limit for $N=2,L=20,U=4,\theta=-\pi/2$. $\alpha=0.1$. Left panel shows the particle density $\rho(x,t)$, and right panel shows the density imbalance $\Delta N$ between left and right halves of the system. 
While interference between particles lead to different branches of wavefunction pumped toward left and right, the density imbalance  $\Delta N$  is found to increases monotonically.
}
\end{figure}

\end{widetext}

\end{document}